\documentclass[10pt,journal]{IEEEtran}
\usepackage{amsmath,amsfonts}
\usepackage{array}
\usepackage{subfigure}
\usepackage{textcomp}
\usepackage{stfloats}
\usepackage{url}
\usepackage{verbatim}
\usepackage{graphicx}
\usepackage{cite}
\usepackage{makecell}
\usepackage{qcircuit}
\usepackage{braket}
\usepackage{array}
\usepackage{slashed}
\usepackage{tikz}
\usepackage{stfloats}
\usepackage{makecell}
\usepackage{multirow}
\usepackage{enumitem}
\usepackage{url}
\usepackage{verbatim}
\usepackage{graphicx}
\usepackage{tabularx}
\usepackage{algpseudocode}
\usepackage{booktabs}
\usepackage{amssymb}
\usepackage{sharina}
\usepackage{float}

\usepackage[ruled,norelsize,vlined,linesnumbered]{algorithm2e}
\makeatletter
\newcommand{\removelatexerror}{\let\@latex@error\@gobble}
\makeatother

\removelatexerror

\begin{document}

\title{Evidential Quantum Vertical Federated Learning}

\author{Hao~Luo, Zhiyuan~Zhai, Qianli~Zhou, Jun~Qi, Yong~Deng, Xin~Wang$^{\ast}$
%
\thanks{Hao Luo, Zhiyuan Zhai, and Xin Wang are with the College of Future Information Technology, Fudan University, Shanghai 200438, China.}
\thanks{Jun Qi is with the School of Electrical and Computer Engineering, Georgia Institute of Technology,
	Atlanta, GA 30332, USA.}
\thanks{Qianli Zhou is with the School of Electronics and Information,	Northwestern Polytechnical University, Xi’an 710072, China.}
\thanks{Yong Deng is with the Institute of Fundamental and Frontier Science, University of Electronic Science and Technology of China, Chengdu 610054, China and also with the School of Medicine, Vanderbilt University, Nashville 37240, USA.}

}

\markboth{Journal of \LaTeX\ Class Files,~Vol.~14, No.~8, August~2021}%
{Shell \MakeLowercase{\emph{et al.}}: A Sample Article Using IEEEtran.cls for IEEE Journals}


\maketitle

\begin{abstract}
Quantum federated learning (QFL) has recently emerged as a promising paradigm for privacy-preserving collaborative learning, yet most existing studies focus on horizontal federated learning and ignore the vertical federated learning (VFL), where parties hold complementary features of aligned samples. In this work, we propose Evidential Quantum Vertical Federated Learning (eviQVFL), a VFL-tailored QFL framework that employs a hybrid classical-quantum architecture for party-side feature processing, mapping local features into a quantum state. To preserve privacy and avoid information loss, party-side output states are directly transmitted to the server via quantum teleportation, and the server fuses the received quantum states with a non-parametric evidential fusion circuit grounded in evidence theory, followed by measurement-based inference.  Extensive simulations on image classification and other real-world datasets demonstrate that eviQVFL consistently achieves higher classification accuracy than other classical and quantum baselines under comparable parameter budgets. Both empirical observations and theoretical analysis indicate that eviQVFL achieve less approximation error with limited quantum resources, while maintaining training stability and offering stronger feature privacy.

\end{abstract}


\section{Introduction}
\IEEEPARstart{Q}{uantum} computing offers a fundamentally different model of computation grounded in superposition, entanglement, and interference of quantum mechanics~\cite{nielsen2010quantum, liu2025toward}. Using quantum bits (qubits) as the unit of computation and quantum circuits as the modeling paradigm, quantum algorithms have been proposed and are regarded as achieving quantum supremacy over classical counterparts; e.g., factoring~\cite{shor1999polynomial} and unstructured search~\cite{grover1996fast}. However, constrained by the current state of quantum hardware, we remain in the noisy intermediate-scale quantum (NISQ) era~\cite{preskillNISQ2018}, where only noisy quantum processors with limited qubit counts, circuit depths, and coherence times are available~\cite{shanmugam2023unrolling,xue2025generalized}. In this context, the field of quantum machine learning (QML) has emerged~\cite{biamonte2017quantum, cerezo2022challenges}, with the aim of harnessing NISQ quantum devices and the unique mechanisms of quantum computing to introduce new insights into machine learning~\cite{Yu2022a,qi2023theoretical}. A canonical framework is the quantum neural network (QNN)~\cite{cerezoVQA2021}, which is composed by a variational quantum circuit (VQC) with trainable rotation angle parameters. QNN enables mapping data features to quantum states of the exponentially large Hilbert space and carries out the inference by evolving these states with VQCs. As in classical neural networks, parameter updates in QNN are performed via gradient-based optimization executed by a classical computer. With fewer parameters and stronger representational capacity, QNNs are considered to achieve advantageous applications based on NISQ devices. Based on QNNs, numerous quantum algorithms such as  classification~\cite{li2022recent}, time-series analysis~\cite{li2024quantum}, and generative learning~\cite{tian2023recent}, etc., have been proposed and shown to hold the potential of exhibiting quantum advantages over their classical counterparts.

These successful precedents have motivated researchers to deploy quantum computing across a broader range of machine learning subfields. Among these, federated learning (FL) is an important branch~\cite{konevcny2016federated}. It is a training framework where a central server and multiple parties collaboratively train a model while preserving the privacy of local data. The integration of quantum computing and FL leads to quantum federated learning (QFL), aiming to strengthen privacy and security, increase computational efficiency, achieve superior model performance, and support quantum-native data~\cite{qiao2024transitioning,ren2025toward}. In recent work, a wide variety of QFL frameworks were proposed. In 2021, Chen \emph{et al}. first put forward the concept of federated QML and proposed a framework for federated training on hybrid quantum-classical classifiers~\cite{chen2021federated}.  QuantumFed is another framework~\cite{xia2021quantumfed}, where multiple quantum nodes conduct collaborately learning using local quantum data, expanding the application scope for FL. Also, blind quantum computing was integrated in QFL to provide privacy security under gradient attack~\cite{li2021quantum}. Additionally, researchers are exploring diverse QNN architectures. Chehimi \emph{et al}. employed quantum convolutional neural networks (QCNN) to classify quantum data derived from quantum many-body systems~\cite{chehimi2022quantum}. And   Qi  put forth the  federated quantum natural gradient descent (FQNGD) algorithm that admits  fewer training iterations and communication costs~\cite{qi2024federated}.

However, the current QFL studies adopt the horizontal FL (HFL) setting~\cite{yang2019federated}, where parties hold disjoint sets of samples over a shared feature space (“same features, different samples”). In HFL, aggregating model parameters, gradients, and local updates is relatively straightforward since parties share the same local model. Thus, HFL dominated the early explorations. In contrast, many real-world problems involve collaboration across different organizations. For example, banks may evaluate customers’ credit risk based on their e-commerce consumption patterns and social media usage, where the data are vertically partitioned across different partiese~\cite{cha2021implementing, zheng2020vertical,cai2022privacy}. In vertical FL (VFL), multiple parties observe the same set of entities, but each holds a distinct, complementary subset of features (“same samples, different features”)~\cite{liu2024vertical}. The objective of VFL is to train a globally joint model spanning all parties and the server, which is fundamentally different from HFL. Deploying quantum techniques to the VFL setting is technically promising and also challenging. As guaranteed by quantum mechanics, the no-cloning theorem and secure quantum communication can inherently ensure the data privacy during the collaborate training process. Meanwhile, a possible bottleneck is the design of the quantum circuit for cross-party information fusion, which should efficiently aggregate local quantum states from different parties without exposing local features. Especially, when fusing a large number of qubits, the training of the prior VQC-based models are highly prone to encountering the barren plateau problem, where the gradients approach zero and the model cannot be effectively trained~\cite{mcclean2018barren,cerezo2021cost}. In this paper, we propose Evidential Quantum Vertical Federated Learning (eviQVFL) framework for solving the VFL scenario.

The evidential part of our framework is grounded in evidence theory, also known as the Dempster-Shafer theory, which is a power set extension of classical probability theory~\cite{dempsterProbabilities1967, shafer1976mathematical}. By assigning the belief mass to $2^n$ power set elements rather than $n$ singletons, evidence theory enables more precise uncertainty representation and information fusion. Over the past period, it has been extensively applied in information fusion~\cite{yang2013evidential}, decision making~\cite{mi2015belief}, and pattern classification~\cite{huang2023combination}. Concurrently, evidence theory has been integrated with deep learning to quantify uncertainty of neural network models, giving birth to the evidential deep learning (EDL) which is applied in tasks such as classification~\cite{sensoy2018evidential} and regression~\cite{amini2020deep}. However, current EDL formulations employ only a simplified power set rather than the full power set, thereby failing to exploit the complete expressive capacity of evidence theory. The primary reason is that the number of elements in the power set grows exponentially, rendering full-scale computation prohibitively expensive. Coincidentally, an $n$-qubit quantum system has $2^n$ computational basis states, whose amplitudes can be mapped to the elements of a power set~\cite{panQuantum2022,zhou2023bfqc}. Prior work shows that by encoding evidence into quantum superposition, one can leverage quantum circuits to conduct evidence evolution efficiently in Hilbert space~\cite{luo_attribute_2024,zhou2024transferable}, whereas classical algorithms would incur exponentially growing computational costs. Therefore, evidence theory is naturally compatible with QNNs: On the one hand, quantum superposition overcomes the prohibitive computational overhead; on the other hand, evidential semantics help guide the design of quantum circuits and enable more precise uncertainty operations as the power set support. In eviQVFL framework, we endow quantum states with evidential semantics, use QNNs to realize local evidence evolution, and finally perform evidence fusion on the server-side quantum circuits.

The main contributions of our work are summarized as follows. We propose {Evidential Quantum Vertical Federated Learning (eviQVFL)} for the VFL scenario, which adopts a {hybrid classical-quantum} architecture for local feature extraction, transforming its input feature vector into an output quantum state. To enable collaboration without exposing local features, the framework transmits party-side output state to the server with {privacy protection via quantum teleportation}, after which the server deploys an {evidential quantum circuit} to fuse multi-party information and conducts {measurement-based inference}. The design of the non-parametric evidential fusion layer is informed by the evidence-theoretic principles. The eviQVFL framework is shown to be both effective and trainable in practice, consistently improving classification accuracy over classical and quantum baselines while also alleviating barren plateaus during VQC training.

The rest of this paper is outlined as follows. Section II introduces the preliminaries of VFL, quantum computing, and evidence theory. Section III details each building block of the proposed eviQVFL framework. In Section IV, we conduct comprehensive experiments for multiple classification tasks and also presents the theoretical analysis for the  eviQVFL framework. The conclusion is finally provided in Section V.

\section{Preliminaries}
\subsection{Classical vertical federated learning}\label{sec:ver}
\begin{figure}[htbp]
	\centering
	\begin{minipage}{0.48\linewidth}
		\centerline{\includegraphics[width=4cm]{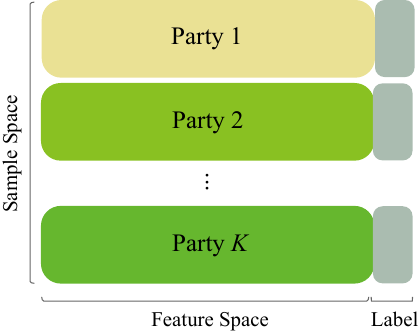}}
		\centerline{(a) HFL}
	\end{minipage}
	\hfill
	\begin{minipage}{.48\linewidth}
		\centerline{\includegraphics[width=4cm]{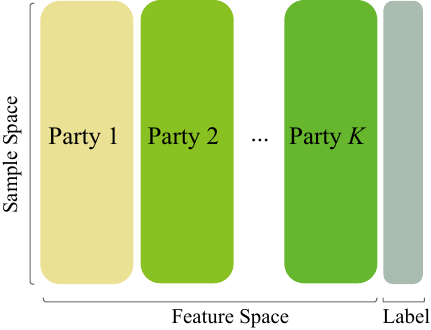}}
		\centerline{(b) VFL}
	\end{minipage}
	\caption{The comparison between HFL and VFL. }
	\label{fig:fl_comparison}
\end{figure}

Federated Learning (FL) enables multiple parties to collaboratively train machine learning models without centralizing raw data. 
According to how data are partitioned among different parties, FL is commonly instantiated as \emph{Horizontal FL (HFL)} and \emph{Vertical FL (VFL)} (see Fig.~\ref{fig:fl_comparison}). Unlike HFL, different parties share the same sample but hold different local features in VFL. Besides, the labels are only available at the server side. Thus, VFL aims to train a joint model that is composed by all the party-side local models and the server-side model, while preserving the privacy of local feature data.

A VFL system aims to collaboratively train a joint model using a dataset
$\mathcal{D} \triangleq \{(\mathbf{x}_i, y_i)\}_{i=1}^{N}$ with $N$ samples. As VFL assumes that data are partitioned by feature space, and each feature vector $\mathbf{x}_i \in \mathbb{R}^{ d}$ in $\mathcal{D}$ is distributed among $K$ parties $\{\mathbf{x}_{i,k}\in \mathbb{R}^{ d_k}\}_{k=1}^{K}$, where $\mathbf{x}_{i,k}$ is the feature held by party $k$ for $k\in\{1,\ldots,K\}$. Thus each party $k$ has dataset $\mathcal{D}_k \triangleq \{\mathbf{x}_{i,k}\}_{i=1}^{N}$; and the server holds the label information $\yy_i$. The joint VFL model  is divided into local modules $\mathcal{G}_k$ parameterized by $\boldsymbol{\theta}_k$, $k\in\{1,\ldots,K\}$ at the party side, and a global module $\mathcal{F}$ parameterized by $\boldsymbol{\gamma}$, which is only accessible by the server. During the inference stage, each party side generates the intermediate result $ \gg_{i,k}=\mathcal{G}_k(\xx_{i,k},\theta_k)$ by the local module and sends it to the server. Then, the global module aggregates all the intermediate results with the global module to produce the final output $\hat{\yy}_i=\mathcal{F}(\gg_{i,1},\dots,\gg_{i,K},(\boldsymbol{\gamma}))$, where the global module $\mathcal{F}$ can be non-trainable, and parameters $\boldsymbol{\gamma}$ no longer exis. After all, the training objective of VFL is to minimize the loss defined by:
\begin{equation}
    \min_{\boldsymbol{\theta}_1,\dots,\boldsymbol{\theta}_K,(\boldsymbol{\gamma})}\frac{1}{N}\sum_{i=1}^{N}
\mathcal{L}\!\left(
  \hat{\yy}_i, \yy_i
\right),
\end{equation}
where $\mathcal{L}$ denotes the  loss function.

\subsection{Quantum computing}
Quantum computing is an emerging paradigm that leverages the principles of quantum mechanics to process information. Unlike classical computing, which uses bits that represent either 0 or 1, quantum computing utilizes quantum bits (qubits) that can exist in a \emph{superposition} of states. Formally, a pure single qubit state is described by linear combination under the normalization constraint: 
\begin{equation}
	\ket{\psi} = \alpha \ket{0} + \beta \ket{1}, \quad \mathrm{with}\;\; \alpha, \beta \in \mathbb{C}, \ |\alpha|^2 + |\beta|^2 = 1,
	\label{eq:qubit_superposition}
\end{equation}
where $\ket{0}, \ket{1}$ are basis states with vector correspondence as $(1,\ 0)^H$ and $(0,\ 1)^H $ respectively. And $\ket{\cdot}$ is the Dirac notation used to represent the state. All the possible pure states $\ket{\psi}$ expand to a two-dimensional Hilbert space $\mathcal{H}_2$. For multiple qubits, the whole state can be constructed as tensor product $ \ket{\psi}=\ket{\psi_1\dots \psi_n}=\ket{\psi_1}\otimes\dots\otimes\ket{\psi_n}$, falling in exponential large Hilbert space $\mathcal{H}_{2^n}$.

Due to the normalization constraint, the quantum operations are restricted as unitary transformations $\ket{\psi}\rightarrow \textbf{U}\ket{\psi}$, where $\textbf{U}$ is a unitary. To better describe sophisticated quantum computing processes, quantum circuits are introduced, denoting a wire as a qubit.  Table~\ref{tab:quantum_gate} summarize most commonly used quantum gates with their corresponding forms on quantum circuits and unitary representations.
\begin{table}[htbp]
	\centering
	\caption{Common quantum gates.}
	
	\begin{tabular}{ccc}
		\hline
		Quantum Gates & Quantum Circuits & Unitaries\\
		\hline
		\rule{0pt}{5ex}Pauli-X gate & $\Qcircuit @C=1em @R=.7em {
			& \gate{X} & \qw
		}$  & $X = \begin{pmatrix}0 & 1 \\1 & 0\end{pmatrix}$ \\ 
		\rule{0pt}{5ex}Pauli-Y gate & $\Qcircuit @C=1em @R=.7em {
			& \gate{Y} & \qw
		}$ & $Y=\begin{pmatrix}0 & -i \\i & 0\end{pmatrix}$  \\ 
		\rule{0pt}{5ex}Pauli-Z gate & $\Qcircuit @C=1em @R=.7em {
			& \gate{Z} &   \qw
		}$ & $Z = \begin{pmatrix}1 & 0 \\0 & -1\end{pmatrix}$ \\ 
		\rule{0pt}{5ex}Hadamard gate & $\Qcircuit @C=1em @R=.7em {
			& \gate{H} & \qw
		}$ & $ H = \begin{pmatrix}1 & 1 \\1 & -1\end{pmatrix}/\sqrt{2}$ \\ 
		\rule{0pt}{5ex}CNOT gate & $\Qcircuit @C=1em @R=.7em {
			& \ctrl{1} &  \qw \\
			& \targ &  \qw
		}$  & $ \mathrm{CNOT}= \begin{pmatrix}I & 0 \\0 & X \end{pmatrix}$ \\
		\rule{0pt}{5ex}Rotation-X gate & $\Qcircuit @C=1em @R=.7em {
			& \gate{R_x(\theta)} & \qw
		}$ &$R_x(\theta)=\exp(-i\theta X/2)$ \\
		 \rule{0pt}{5ex}Rotation-Y gate & $\Qcircuit @C=1em @R=.7em {
		 	& \gate{R_y(\theta)} & \qw
		 }$ &$R_y(\theta)=\exp(-i\theta Y/2)$ \\
		 \rule{0pt}{5ex}Rotation-Z gate & $\Qcircuit @C=1em @R=.7em {
		 	& \gate{R_z(\theta)} & \qw
		 }$ &$R_x(\theta)=\exp(-i\theta Z/2)$ \\
		\hline
		
	\end{tabular}
	\label{tab:quantum_gate}
\end{table}

\begin{figure}[htbp]
	\centering
	\mbox{\Qcircuit @C=1em @R=1.5em {
	\lstick{\ket{0}}& \qw &	\gate{H} & \qw  & \ctrl{1} & \barrier[-0.9em]{1} \qw  & \qw \\
	\lstick{\ket{0}}& \qw & \qw	& \qw  &\targ &  \qw & \qw\\
	&&&&&\mbox{$\ket{\Psi^+}$}&
		}
	}
	\caption{The quantum circuit of preparing EPR pair.}
	\label{fig:epr_generate}
\end{figure}
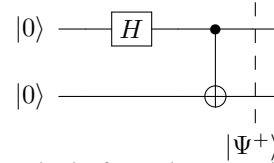

Another key concept in quantum computing is \emph{quantum entanglement}, wherein two (or more) qubits become correlated in ways that have no classical analog. A prime example is the Einstein–Podolsky–Rosen (EPR) pair, also known as a Bell state:
\begin{equation}\label{eq:epr}
	\ket{\Psi^+}=(\ket{00}+\ket{11})/\sqrt{2},
\end{equation}
which can be easily prepared by operations of Hadamard gate and CNOT gate as depicted in Fig.~\ref{fig:epr_generate}.

For a EPR pair, if one qubit is measured and found to be in $\ket{0}$, the other qubit is guaranteed to be in $\ket{0}$ as well; likewise, if the first qubit is observed in $\ket{1}$, the second qubit will also be $\ket{1}$. This perfect correlation, revealing the entanglement between both qubits, holds regardless of the distance, serving as essential resources in various quantum communication protocols such as quantum teleportation that will be mentioned later.

\subsection{Evidence theory and its quantum equivalence}

Evidence theory, also referred to as Dempster–Shafer theory~\cite{dempsterProbabilities1967,shafer1976mathematical}, offers a generalized framework for reasoning under uncertainty by assigning belief to subsets of a hypothesis space, known as the power set rather than individual elements in conventional probability theory. 

Let $\Omega = \{\omega_1,\omega_2,\dots,\omega_n\}$ denote the \emph{frame of discernment} and $2^\Omega=\{\varnothing,\{\omega_1\},\{\omega_2\},\{\omega_1,\omega_2\},\dots,\Omega\}$ its power set. A key concept in evidence theory is the \emph{basic belief assignment} (BBA), denoted by a mass function $m(\cdot)$, which allocates belief masses to the power set:
\begin{equation}
	m: 2^{\Omega} \;\rightarrow\; \mathbb{R}, 
	\quad \sum_{F\subseteq \Omega} m(F)\;=\;1,
	\label{eq:BBA_def}
\end{equation}
where $m(F)$ indicates the degree of belief specifically committed to the proposition $F \subseteq \Omega$. Besides, the plausibility function provides an upper bound on the support for a hypothesis. For any $F\in\Omega$, it is defined as
\begin{equation}
	\mathrm{Pl}(F)=\sum_{\substack{G\cap F\neq \varnothing\\G\in 2^\Omega}}m(G),
	\label{eq:plausibility_def}
\end{equation}
which sums $2^{n-1}$ the belief masses that do not rule out $F$, representing the maximum potential support for $F$.

When dealing with multiple independent pieces of evidence, evidence theory provides a branch of combination rules to fuse them. One of the most fundamental rules is the \emph{conjunctive combination rule (CCR)}. Suppose we have $K$ pieces of evidence, each represented by a mass function $m_k(\cdot)$ for
$k = 1,2,\dots,K$. The CCR aggregates them into a fused mass function $m(\cdot)$ defined by:
\begin{equation}
	m(F) \;=\; \sum_{\substack{F_1 \cap F_2 \cap \cdots \cap F_K = F}} 
	\prod_{k=1}^{K} m_k\bigl(F_k\bigr),
	\label{eq:ccr}
\end{equation}
where each $F_k$ is an element in $2^\Omega$, and $m(F)$ captures the combined evidence supporting $F$ based on the intersection of all $F_k$. This conjunctive rule emphasizes the overlap among different sources, reinforcing the common beliefs.

While evidence theory is powerful, handling the $2^n$ elements in the power set leads to exponential computational complexity. Interestingly, the information in the power set can be naturally mapped to the superposition states of a quantum system~\cite{panQuantum2022, zhou2023bfqc, luo_attribute_2024, zhou2024transferable}, where quantum entanglement enables simultaneous operations on all elements, achieving exponential acceleration.

Concretely, one can encode classical evidence of an $n$-element frame of discernment $\Omega$ into a \emph{quantum evidence state} by mapping the BBA values, augmented with random phases, to the amplitudes of an $n$-qubit superposition. Let $\{F_x: x\in\{0,1\}^n\}$ denote the subsets corresponding to the computational basis. For example, when $n=2$ and $\Omega=\{\omega_1,\omega_2\}$, one may define $F_{00}=\varnothing$, $F_{01}=\{\omega_2\}$, $F_{10}=\{\omega_1\}$, and $F_{11}=\{\omega_1,\omega_2\}$. Then, the mapped $n$-qubit quantum state is expressed as
\begin{equation}
	\ket{\psi} = \sum_{x_1,\dots,x_n\in\{0,1\}} \sqrt{m(F_{x_1\dots x_n})}\ \mathrm{exp}(i\beta_{x_1\dots x_n})\ \ket{x_1\dots x_n},
	\label{eq:quantum_state}
\end{equation}
where $\beta$ denotes the random phase. According to (\ref{eq:quantum_state}), each element in $\Omega$ corresponds to a single qubit. Thus, each element $F_{x_1\dots x_n}$ in $2^\Omega$, where evidence theory is defined above, corresponds to a ground state $\ket{x_1\dots x_n}$. Furthermore, given an arbitrary $n$-qubit quantum state, it can conversely symbolize a piece of evidence according to the mapping in (\ref{eq:quantum_state}), once we assign the meaning of elements in a frame of discernment $\Omega$ to qubits. This offers the quantum system interpretability through the evidence semantics.

\section{The eviQVFL framework}
In this section, we extend the classical VFL model introduced in~\ref{sec:ver} to a fully quantum framework. We begin by presenting the overall architecture of the framework in~\ref{subsec:qvfl_overall}. The design and implementation details of each component are presented in the following subsections~\ref{sec:party}~\ref{sec:teleport}~\ref{sec:agg}. 

\subsection{The overall quantum VFL framework}
\label{subsec:qvfl_overall}
We consider a classification task of $K$ parties collaboratively learning from $N$ aligned samples, where a privacy-preserving entity alignment is assumed to have been performed beforehand. For sample $i\in\{1,\ldots,N\}$, party $k\in\{1,\ldots,K\}$ owns a local feature block $\mathbf{x}_{i,k}\in\mathbb{R}^{d_k}$, and the server holds the one-hot label $\mathbf{y}_i\in\mathbb{R}^{C}$, where $C$ denotes the number of classification types.

Each party implements a hybrid quantum–classical model $\mathcal{G}_k(\cdot;\boldsymbol{\theta}_k)$ on $n_k$ qubits, which encodes local feature data to a quantum state expressed as:
\begin{equation}
\label{eq:local_state}
\ket{\psi_{i,k}}
  = \mathcal{G}_k(\mathbf{x}_{i,k};\boldsymbol{\theta}_k).
\end{equation}
Here, $\boldsymbol{\theta}_k$ denotes local model parameters for both classical and quantum models.

Once the local output is prepared, the quantum state $\ket{\psi_{i,k}}$ is transmitted to the server via \emph{quantum teleportation}, ensuring that the information is conveyed without physically moving the qubits and without revealing the underlying classical features. Notably, directly transmitting the quantum state rather than its classically measured outcomes avoids the information loss inherent in measurement, preserving the full quantum description of the local encoding. Moreover, quantum-state transmission inherently protects against raw data leakage, as no classical feature values might be exposed, and the framework also avoids the number of quantum–classical conversions, thereby improving both training and inference efficiency.

After receiving the states $\{\ket{\psi_{i,1}},\ldots,\ket{\psi_{i,K}}\}$ from all parties, the server applies a global fusion module \(\mathcal{F}(\cdot;(\boldsymbol{\gamma}))\) acting on the joint state in the composite Hilbert space:
\begin{equation}
\label{eq:fusion}
\ket{\phi_i}
= \mathcal{F}\!\left(\ket{\psi_{i,1}}\otimes\cdots\otimes \ket{\psi_{i,K}};(\boldsymbol{\gamma})\right),
\end{equation}
where $(\boldsymbol{\gamma})$ denotes the optional parameters in the fusion model. After fusion, the output state $\ket{\phi_i}$ contains the fused representation for all the parties.

The fused state $\ket{\phi_i}$ is then measured on a readout subsystem with a projective measurement $\{\Pi_c\}_{c=1}^C$, yielding the class probabilities with the softmax function:
\begin{equation}
\label{eq:prob}
\hat{y}_i(c)=\mathrm{softmax}_c\Big(\bra{\phi_i}\Pi_c\ket{\phi_i}\Big), 
\end{equation}
where $\bra{\phi_i}$ denotes the conjugate transpose of $\ket{\phi_i}$; and $\Pi_c$ denotes the defined measurement operator corresponding to the classification type $c$, mapping the quantum state $\ket{\phi_i}$ to a classical predictive support probability $\hat{y}_i(c)$ for the type $c$. When applying the quantum VFL
model during the inference, the classification result should be $\arg\max_{c} \hat{y}_i(c)$.

\begin{figure}[htbp]
	\centering
	\includegraphics[width=7cm]{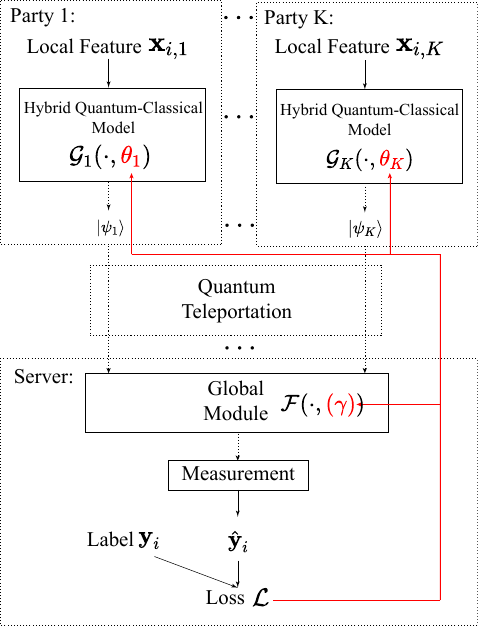}
	\caption{Overall architecture of the proposed quantum vertical VFL framework. Solid lines denote classical communication, dashed lines denote quantum communication. Black arrows indicate the forward-propagation flow of quantum states or classical signals, while red arrows indicate the classical update according to the gradients.}
	\label{fig:qfl}
\end{figure}

Below, we discuss the learning process for the quantum VFL model. The learning objective is defined as
\begin{equation}
\label{eq:objective}
\min_{\{\boldsymbol{\theta}_k\}_{k=1}^K,\,(\boldsymbol{\gamma})}
\frac{1}{N}\sum_{i=1}^N \mathcal{L}\!\left(\hat{\yy}_i,\,\mathbf{y}_i\right)
,
\
\mathbf{\hat{y}}_i=[\hat{y}_i(1),\ldots,\hat{y}_i(C)],
\end{equation}
where $\mathcal{L}$ denotes the loss function. In this paper, the cross-entropy loss is adopted for classification tasks. According to the above modeling, the training of local and global parameters transforms to the training of the joint model for a smaller loss. Thus, gradient descent is applied for updating parameters, which are implemented via a classical optimizer. In this way, only quantum states and classical gradients are exchanged; raw features $\mathbf{x}_{i,k}$ and labels $\textbf{y}_i$ remain entirely private, as illustrated in Fig.~\ref{fig:qfl}. Based on the quantum VFL model, we will introduce the eviQVFL framework in the subsections below, where the hybrid quantum-classical model, quantum teleportation, and global module will be discussed in detail, respectively.

\subsection{Party-side feature processing}\label{sec:party}
In a general setting, a hybrid quantum-classical model is composed of: a classical neural network, a quantum encoding layer, and a variational quantum circuit (VQC). In eviQVFL, the tensor-train Network and Variational Quantum Circuit (TTN–VQC) architecture is utilized as the hybrid quantum-classical model at the the party side. The whole pipeline is as follows. First, the classical TTN maps the high-dimensional local feature into a low-dimensional one; Second, the initial quantum state can be prepared by encoding the low-dimensional feature; Finally, the quantum state goes through a parameterized unitary operator specified by a VQC, which outputs a quantum state $\ket{\psi_{i,k}}$. 

It is worth noting that the TTN part and the encoding part are not mandatory in all cases. Their inclusion depends on the nature of the local features in the target application. When the input $\mathbf{x}_{i,k}$ is a high-dimensional classical feature vector—such as those extracted from images, audio signals, or other rich sensory data—the complete pipeline is typically required. If the input consists of low-dimensional classical features, for instance, tabular attributes like credit card transactions or medical examination indicators, the dimensionality reduction stage can be omitted, and the quantum embedding layer can directly map the features to qubits. 
In scenarios where the input is already given in the form of a quantum state, such as the output of a quantum sensor or another quantum subsystem, the process can bypass both the classical neural network and embedding, feeding the state directly into the VQC for further processing. 
Without loss of generality, we denote the parameters for TTN and VQC as $\boldsymbol{\theta}_k^{\mathrm{ttn}}$ and $\boldsymbol{\theta}_k^{\mathrm{vqc}}$, both of which compose the parameter $\boldsymbol{\theta}_k$ for the $k$-th party side.

\paragraph{Dimension reduction by TTN}
Although many classical models, such as principal component analysis (PCA) and dense neural networks (DNN), can serve as a dimension reduction layer, we apply the tensor-train network (TTN) in the paper since it enables fewer parameter counts and end-to-end trainability without sacrificing expressive power~\cite{qi2023qtn}. 

TTN is based on the TT decomposition~\cite{oseledets2011tensor} and has been commonly employed in machine learning tasks like speech processing~\cite{qi2020tensor} and computer vision~\cite{yang2017tensor}. In detail, TTN factorizes both data and trainable operators into local cores with small bond dimensions.  The input vector $\mathbf{x}_{i,k}\in\mathbb{R}^{d_k}$ is first reshaped into a $L$-order tensor $\mathcal{X}\in \mathbb{R}^{P_1\times P_2\times\dots \times P_L}$, where $P_l$ denotes the dimension for the $l$-th order and satisfies $\prod_{l}P_l=d_k$. Tensor $\mathcal{X}$ is further decomposed into a TT core form:
\begin{equation}
    \mathcal{X}(p_1,\dots,p_L)= \prod_{l=1}^L\mathcal{X}^{(l)}(p_l),
\end{equation}
where $\mathcal{X}^{(l)}\in\mathbb{R}^{R^{\mathcal{X}}_{l-1}\times P_l\times R^{\mathcal{X}}_{l}}$ denotes the TT cores and $R^{\mathcal{X}}_{l}$ denotes the ranks for the TT bond. The boundary ranks $R^{\mathcal{X}}_{0}=R^{\mathcal{X}}_{L}=1$, and $\mathcal{X}^{(l)}(p_l)\in\mathbb{R}^{R^{\mathcal{X}}_{l-1}\times R^{\mathcal{X}}_{l}}$ represents a 2-dimension matrix sliced from $\mathcal{X}^{(l)}$. 

\begin{figure}[htbp]
	\centering
	\includegraphics[width=8cm]{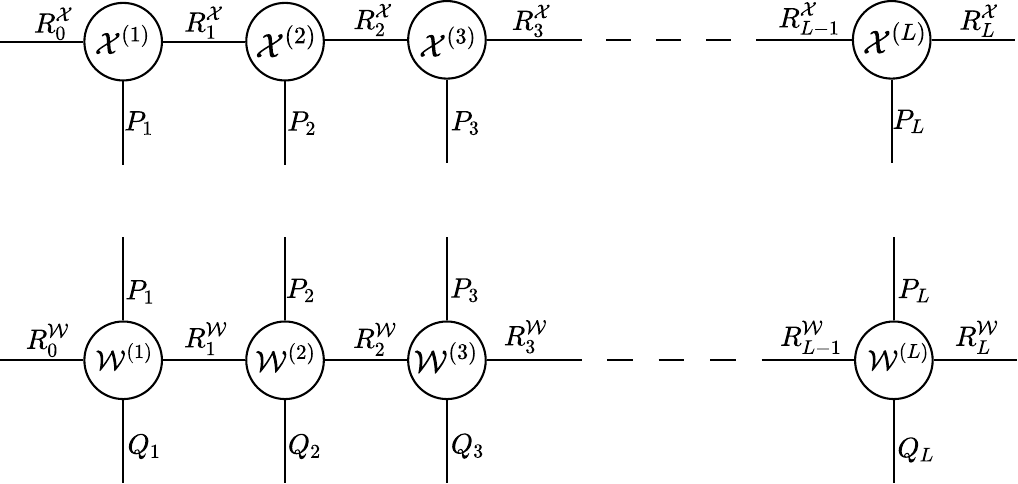}
\caption{Tensor Train Network (TTN) acting on an input represented in Tensor Train (TT) form. The top row shows TT cores $\mathcal{X}^{(l)} \in \mathbb{R}^{R^{\mathcal{X}}_{l-1} \times P_l \times R^{\mathcal{X}}_{l}}$ for $l=1,\dots,L$. The bottom row shows the TT-operator cores $\mathcal{W}^{(l)} \in \mathbb{R}^{R^{\mathcal{W}}_{l-1} \times P_l \times Q_l \times R^{\mathcal{W}}_{l}}$. Each pair of corresponding cores is contracted along the shared input mode $P_l$, producing an output tensor of size $Q_1 \times \cdots \times Q_L$. Vertical legs denote physical modes ($P_l$, $Q_l$); horizontal legs denote TT bond dimensions $R^{\mathcal{X}}_{l}$ and $R^{\mathcal{W}}_{l}$ (with boundary ranks $R^{\mathcal{X}}_{0}=R^{\mathcal{X}}_{L}=R^{\mathcal{W}}_{0}=R^{\mathcal{W}}_{L}=1$).}
	\label{fig:ttn}
\end{figure}

To map $\mathcal{X}$ to an output tensor $ \tilde{\mathcal{X}}$, we introduce a trainable TTN operator with cores $\mathcal{W}^{(l)}\in\mathbb{R}^{R^{\mathcal{W}}_{l-1}\times P_l\times Q_l\times R^{\mathcal{W}}_{l}}$ where  $R^{\mathcal{W}}_{l}$ is the ranks for the TT bond. The ranks can be set artificially, which only needs to guarantee that boundary ranks satisfy
$R^{\mathcal{W}}_{0}=R^{\mathcal{W}}_{L}=1$. TTN greatly save the sacle of parameter numbers since the number of total parameters required for constructing all the cores is the summation $\sum_{l=1}^LR_{l-1}^{\mathcal{W}}P_lQ_lR_{l}^{\mathcal{W}}$, rather than the product $\prod_{l=1}^LP_l\prod_{l=1}^LQ_l$ in DNN. The action of the TTN on the TT input is the contraction over the shared modes $P_l$. We define $\tilde{\mathcal{X}}^{(l)}$ as the TT cores for the output, thus the contraction can be expressed as:
\begin{equation}
\begin{aligned}
        \tilde{\mathcal{X}}(q_1,\dots,q_L)&=\prod_{l=1}^L\left( \tilde{\mathcal{X}}^{(l)}(q_l) \right)\\
        &=\prod_{l=1}^L\left( \sum_{p_l=1}^{P_l} \mathcal{X}^{(l)}(p_l) \otimes \mathcal{W}^{(l)}(p_l, q_l) \right),
\end{aligned}
\end{equation}
where $\tilde{\mathcal{X}}^{(l)}\in\mathbb{R}^{R^{\mathcal{X}}_{l-1}\times R^{\mathcal{X}}_{l}\times R^{\mathcal{W}}_{l-1}\times R^{\mathcal{W}}_{l}\times Q_l} $ is cores of five-order tensors. The output tensor $\tilde{\mathcal{X}}$ is given by contracting $\tilde{\mathcal{X}}^{(l)}$ for $l\in[1,\dots,L]$ over the TT bond dimensions $R^{\mathcal{X}}_{l}$ and $R^{\mathcal{W}}_{l}$. Finally, the tensor $\tilde{\mathcal{X}}$ can be reshaped back to the low-dimensional feature $\tilde{\mathbf{x}}_{i,k}\in \mathbb{R}^{n_k}$, where $n_k=\prod_{l=1}^Lq_l$ corresponds to the number of local qubits available at the party.  The entire process of TTN is presented in Fig.~\ref{fig:ttn}. 

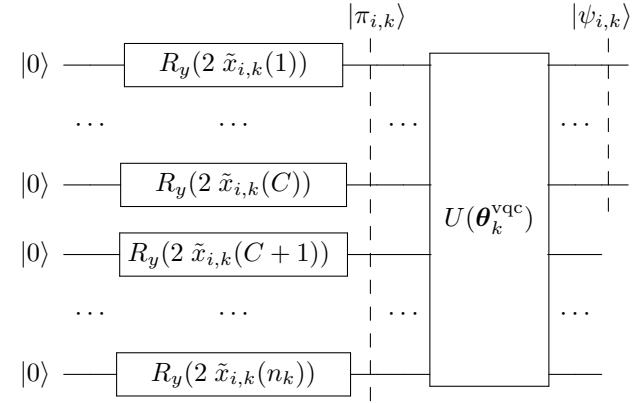
\begin{figure}[htbp]
	\centering
	\mbox{\Qcircuit @C=1em @R=1em {
			&&&\mbox{$\ket{\pi_{i,k}}$}&&&&\mbox{$\ket{\psi_{i,k}}$}& \\
			\lstick{\ket{0}} & \qw & \gate{\quad  R_y(2\;\tilde{x}_{i,k}(1))\quad} \barrier[-0.25em]{5}   & \qw & \qw & \multigate{5}{U(\boldsymbol{\theta}_k^{\mathrm{vqc}})} \barrier[1.25em]{2}& \qw & \qw  & \qw  \\
			\lstick{} & \cdots & \cdots &  & \cdots & \nghost{{U(\boldsymbol{\theta}_k)}} & \cdots &\\
			\lstick{\ket{0}} & \qw & \gate{\quad  R_y(2\;\tilde{x}_{i,k}(C))\quad} & \qw & \qw & \ghost{U(\boldsymbol{\theta}_k^{\mathrm{vqc}})} & \qw & \qw  & \qw \\
			\lstick{\ket{0}} & \qw & \gate{ R_y(2\;\tilde{x}_{i,k}(C+1))\ } & \qw & \qw & \ghost{U(\boldsymbol{\theta}_k^{\mathrm{vqc}})} & \qw & \qw \\
			\lstick{} & \cdots & \cdots &  & \cdots & \nghost{U(\boldsymbol{\theta}_k)} & \cdots & \\
			\lstick{\ket{0}} & \qw & \gate{\quad  R_y(2\;\tilde{x}_{i,k}(n_k))\quad} & \qw & \qw & \ghost{\mathbf{U}(\boldsymbol{\theta}_k^{\mathrm{vqc}})} & \qw & \qw 
		}
	}
	\caption{The quantum circuit at the $k$-th party-side. The feature data $\tilde{\mathbf{x}}_{i,k}$ is first encoded into the quantum state $\ket{\pi_{i,k}}$ by rotation-Y gates. Then, operated by the VQC with trainable parameters $\boldsymbol{\theta}_k^{\mathrm{vqc}}$, the party selects first $C$ qubits as the output, whose quantum state $\ket{\psi_{i,k}}$ carries BBA information.}
	\label{fig:party_circuit}
\end{figure}

\paragraph{Angle encoding}
In the hybrid quantum-classical model, classical features are required to be encoded in the quantum system in preparation for VQC processing. In this paper, the $n_k$-dimension feature $\tilde{\mathbf{x}}_{i,k}$ is mapped to a quantum state of $n_k$ qubit via angle encoding by implementing rotation-Y gates. The quantum circuit of the encoding process is presented in Fig.~\ref{fig:party_circuit}. Assume every qubit is initially set as $\ket{0}$, the encoded quantum state will be:
\begin{equation}
	\begin{aligned}
		\ket{\pi_{i,k}} &=\left(\bigotimes_{t=1}^{n_k} R_y(2\;\tilde{x}_{i,k}(t))\right)\ket{0}^{\otimes n_k} \\&=\bigotimes_{t=1}^{n_k} \Bigl(\cos\bigl(\tilde{x}_{i,k}(t)\bigr)\ket{0} \;+\; \sin\bigl(\tilde{x}_{i,k}(t)\bigr)\ket{1}\Bigr).
	\end{aligned}
	\label{eq:angle_encoding}
\end{equation}
Thus the $n_k$-qubit state $\ket{\pi_{i,k}}$ carries the classical feature information.

\paragraph{VQC}
Once encoded, the encoded state $\ket{\pi_{i,k}}$ will be processed by a VQC, mapping to the output quantum state $\ket{\psi_{i,k}}$. VQC provides a powerful framework for constructing parameterized quantum models capable of representing complex functions, serving as a quantum analog of classical neural networks. VQC is nothing but a trainable unitary operator constructed by a quantum circuit with parameterized gates that enables parameter adjustment by back-propagation during the learning process. Denoting the parameters for the $k$-th party as $\boldsymbol{\theta}_k^{\mathrm{vqc}}\in \mathbb{R}^T$, the unitary operation corresponding to the VQC is written as:
\begin{equation}
	\mathbf{U}(\boldsymbol{\theta}_k^{\mathrm{vqc}})=\prod_{t=1}^{T}\Bigl(\mathbf{U}_{k}^t\exp(-i\mathbf{H}_{k}^t{\theta}_k^{\mathrm{vqc}}(t)\Bigr),
	\label{eq:vqc_ansatz}
\end{equation}
which is composed of multiplying $T$ parametrized layers in total. For the $t$-th layer, the operator $\mathbf{U}_{k}^t\exp(-i\mathbf{H}_{k}^t{\theta}_k^{\mathrm{vqc}}(t))$ is composed of a fixed unitary operator and a parametrized rotation operator specified by a Hermitian operator $\mathbf{H}_{k}^t$ together with a trainable parameter ${\theta}_k^{\mathrm{vqc}}(t)$. A common choice for the Hermitian operator $\mathbf{H}_{k}^t$ is using Pauli strings, which leads to the Pauli rotation gates $R_x$, $R_y$, $R_z$ as listed in Table~\ref{tab:quantum_gate}.

\begin{figure}[htbp]
	\centering
	\mbox{\Qcircuit @C=0.25em @R=1em {
			&\qw&\qw&\gate{R_x}&\gate{R_y}&\gate{R_z}&\ctrl{1}&\qw&\qw&\targ&\qw&\qw&\qw  &\gate{R_x}&\gate{R_y}&\gate{R_z}&\ctrl{1}&\qw&\qw&\targ&\qw&\qw&\mbox{\quad $\dots$} \\
			&\qw&\qw&\gate{R_x}&\gate{R_y}&\gate{R_z}&\targ&\ctrl{1}&\qw&\qw&\qw&\qw &\qw&\gate{R_x}&\gate{R_y}&\gate{R_z}&\targ&\ctrl{1}&\qw&\qw&\qw&\mbox{\quad $\dots$}\\
			&\qw&\qw&\gate{R_x}&\gate{R_y}&\gate{R_z}&\qw&\targ&\qw&\qw&\qw&\qw 	&\qw&\gate{R_x}&\gate{R_y}&\gate{R_z}&\qw&\targ&\qw&\qw&\qw&\mbox{\quad $\dots$} \\
			& & &\vdots&\vdots&\vdots& & &\ddots& & & & &\vdots&\vdots&\vdots& & &\ddots& & & \\
			&\qw&\qw&\gate{R_x}&\gate{R_y}&\gate{R_z}&\qw&\qw&\qw&\ctrl{-4}&\qw&\qw &\qw&\gate{R_x}&\gate{R_y}&\gate{R_z}&\qw&\qw&\qw&\ctrl{-4}&\qw&\mbox{\quad $\dots$}  \gategroup{1}{4}{5}{10}{1em}{--} 
		}
	}
	\caption{The quantum circuit of the VQC structure employed.}
	\label{fig:VQC_circuit}
\end{figure}
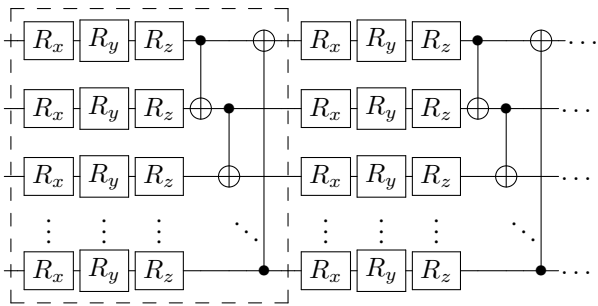

Fig.~\ref{fig:VQC_circuit} shows the VQC structure employed in this paper. The circuit is constructed by repeating the sub-circuit enclosed in the dotted box. For each sub-circuit, $R_x$, $R_y$, and $R_z$ are applied in sequence to adjust the local state of each qubit, and CNOT gates are then applied to introduce entanglement across them. 

Processed by VQC, the input state $\ket{\pi_{i,k}}$ is mapped to $\mathbf{U}(\boldsymbol{\theta}_k^{\mathrm{vqc}})\ket{\pi_{i,k}}$. By adjusting the parameters $\boldsymbol{\theta}_k^{\mathrm{vqc}}$,  the circuit can exploit the exponentially large Hilbert space inherent in quantum systems.  This unique expressive power enables VQCs to approximate intricate functions and potentially achieve advantages in terms of both representational capacity and computational efficiency. 

For the classification task, the set of all the $C$ classification types can be viewed as a frame of discernment of $C$ elements in evidence theory. Assume qubits number $n_k$ is larger than $C$. As suggested by the quantum equivalence of evidence theory, BBA information under the frame of discernment can be carried by only $C$ qubits out of the total of $n_k$ qubits, which means that we can use the quantum state of a subsystem of $C$ qubits as the local output for the party. Without loss of generality, assume that the party selects the first $C$ qubits here. Denote the quantum state of the first $C$ qubits as $\ket{\psi_{i,k}}$, which can be viewed as a quantum evidence state corresponding to a piece of evidence $m_{i,k}$:
\begin{equation}
\begin{aligned}\label{eq:telmass}
      \ket{\psi_{i,k}}=&\sum_{x_1,\dots,x_C\in\{0,1\}}\sqrt{m_{i,k}(F_{x_1\dots x_C})}\\& \quad \mathrm{exp}(i\beta_{k\ x_1\dots x_C})\ \ket{x_1\dots x_C}, 
\end{aligned}
\end{equation}
where $\beta_{k\ x_1\dots x_C}$ is the arbitrary phase that will not influence the final measurement result. Thus, through the state $\ket{\psi_{i,k}}$, the evidence semantics is introduced in the eviQVFL framework. Fig.~\ref{fig:party_circuit} summarizes the whole quantum circuit at the party side, including the encoding part and the VQC part.

\subsection{Quantum teleportation}\label{sec:teleport}

After being processed by VQCs, the parties obtain the quantum states $\ket{\psi_{i,k}}$ carrying their local BBA information, which must be securely transmitted to the server for evidence-theoretic fusion. For the majority of FL frameworks, the communications between party sides and server sides are at risk of data leaks. By contrast, in eviQVFL, we adopt quantum teleportation to achieve this transfer, thereby minimizing the risk of intercepting sensitive raw data and model outputs.

Here, we briefly introduce the principle of quantum teleportation. Quantum teleportation allows a sender to transmit an unknown quantum state to a receiver, which was proposed by Bennett \emph{et al.} \cite{bennett1993teleporting} and first experimentally demonstrated in a photonic quantum system \cite{bouwmeester1997experimental}. In recent years, quantum teleportation has been intensively studied, having progressed from laboratory research to practical applications \cite{hu2023progress}. The technique can be viewed as the key ingredient for applications such as quantum communication \cite{zhou2023towards}, quantum network \cite{simon2017towards}, and quantum internet \cite{wehner2018quantum}, which establish links among distributed quantum computing nodes, providing a technical foundation to realize eviQVFL.

\begin{figure}[htbp]
	\centering
	\footnotesize
	\mbox{\Qcircuit @C=1.3em @R=2em {
			\barrier[0em]{4}&&\barrier[0em]{4}&&\barrier[0em]{4}&&&&&&\mbox{Party Side}&&\\
			\lstick{\ket{\psi}}  & \qw & \ctrl{1} & \qw & \gate{H} & \qw & \meter & \cw  & \cw  & \cw  & \cctrl{2} &&&\\
			 & \qw & \targ & \qw & \qw & \qw & \meter & \cw & \cctrl{1} &&& \\
			 & \qw & \qw & \qw & \qw & \qw & \qw & \qw & \gate{X} & \qw & \gate{Z} & \qw & \rstick{\ket{\psi}}  \inputgroupv{3}{4}{.75em}{1.65em}{\ket{\Psi^+}\ } \gategroup{2}{2}{3}{11}{1.75em}{--}
			 \gategroup{4}{2}{4}{11}{1.5em}{--}\\
			 &&&&&&&&&&\mbox{Server Side}&&\\
			 &\mbox{$\ket{\phi_1}$}&&\mbox{$\ket{\phi_2}$}&&\mbox{$\ket{\phi_3}$}
	}}
	\caption{Standard quantum teleportation circuit. The dotted boxes mark the  Suppose the unknown state $\ket{\psi}$ is to be transferred from the party side to the server side. The party shares an EPR pair $\ket{\Psi^+}$ with the server. After the quantum operations, the party conducts local measurements and sends the two classical bit measurement results to the server, who recovers $\ket{\psi}$ exactly by applying $X$ or $Z$ corrections.}
	\label{fig:teleportation_circuit}
\end{figure}
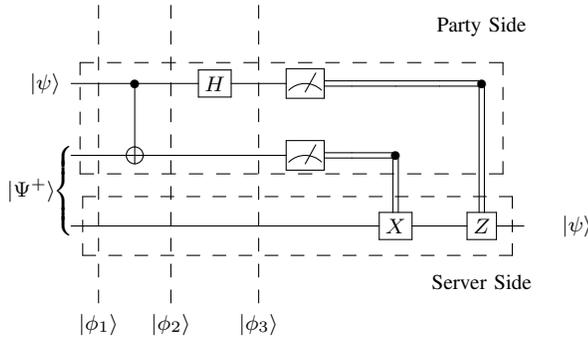

Recalling the eviQVFL framework, the quantum state $\ket{\psi_{i,k}}$ of $C$ qubits is to be transmitted from the $k$-th party side to the server side. As the transmission of multiple qubits can be directly generalized from the single-qubit case \cite{zhao2004experimental, zhang2006experimental}, we only present the pipeline of transmitting a single qubit here. Suppose that the unknown single-qubit state $\ket{\psi}$ is to be transferred. Fig.~\ref{fig:teleportation_circuit} illustrates the process of the standard quantum teleportation.

Initially, the EPR pair $\ket{\Psi^+}$ shown in (\ref{eq:epr}) is required to be prepared and distributed to both sides. Define the state $\ket{\psi}$ as $\alpha\ket{0}+\beta\ket{1}$. Thus, the initial global state $\ket{\phi_1}$ is expressed as:
\begin{equation}
	\begin{aligned}
		\ket{\phi_1}&=\ket{\psi}\otimes\ket{\Psi^+}\\
		&=\frac{1}{\sqrt{2}}\Bigl(\alpha\ket{000}+\alpha\ket{011}+\beta\ket{100}+\beta\ket{111}\Bigr).
	\end{aligned}
\end{equation}

Quantum teleportation starts with the party applying the CNOT gate, which makes the global set to be
\begin{equation}
		\ket{\phi_2}=\frac{1}{\sqrt{2}}\Bigl(\alpha\ket{000}+\alpha\ket{011}+\beta\ket{110}+\beta\ket{101}\Bigr).
\end{equation}

Then, the party applies the Hadamard gate on the qubit to be transferred, which turns the global state to:
\begin{equation}
	\begin{aligned}
		|\phi_3\rangle 
		&= \frac{1}{2} \Bigl(
		\alpha \ket{000} + \alpha \ket{100} + \alpha \ket{011} + \alpha \ket{111} \\
		 & \qquad+ \beta \ket{010} - \beta \ket{110} + \beta \ket{001} - \beta \ket{101} \Bigr)\\
		&=\frac{1}{2} \Bigl( \ket{00}\otimes(\alpha \ket{0}+\beta\ket{1}) + \ket{01}\otimes(\beta \ket{0}+\alpha\ket{1})\\
		&+\ket{10}\otimes(\alpha \ket{0}-\beta\ket{1})+\ket{11}\otimes(-\beta \ket{0}+\alpha\ket{1})\Bigr)
	\end{aligned}
	\label{eq:global_state}
\end{equation}

The next step for the party side is conducting local measurements for both qubits and sending the classical results to the server side. According to (\ref{eq:global_state}), the state of the first two qubits directly determines the state of the third qubit, which is the cause of the existence of the entanglement brought by the EPR pair. After the measurements, the global state collapses into one of four possible outcomes.

In the last step, as shown in Fig.~\ref{fig:teleportation_circuit}, the server will recover the state $\ket{\psi}$ by applying $X$ and $Z$ gates on its local qubit controlled by the received classical measurement results from the party-side, which is shown in Table~\ref{tab:recover}.
\begin{table}[htbp]
	\centering
	\caption{Operations to recover the original state according to the measurement results.}
	\begin{tabular}{ccc}
		\hline
		Measurement Result & Local State & Recover Operation\\
		\hline
		00 & $\alpha \ket{0}+\beta\ket{1}$  & none \\ 
		01 & $\beta \ket{0}+\alpha\ket{1}$  & $X$ gate \\ 
		10 & $\alpha \ket{0}-\beta\ket{1}$  & $Z$ gate \\ 
		11 & $-\beta \ket{0}+\alpha\ket{1}$  & $X$ gate and $Z$ gate \\ 
		\hline
	\end{tabular}
	\label{tab:recover}
\end{table}
In conclusion, transmitting a single-qubit state in teleportation requires two key resources:
\begin{enumerate}
	\item A shared EPR pair, where one qubit is held by the party and the other is held by the server.
	\item Two classical bits of communication from the party to the server.
\end{enumerate}

\subsection{Server-side evidential aggregating}\label{sec:agg}
In eviQVFL, the server receives the output quantum states $\ket{\psi_{i,k}}$ from all the party sides. As (\ref{eq:telmass}) suggests, these states are interpreted as multiple pieces of evidence $m_{i,k}$ containing the predicted belief according to distributed features. In eviQVFL, the CCR aggregating method in (\ref{eq:ccr})  is adopted as the global module to aggregate the belief. Unlike generic VQC ansatz with multiple layers of trainable unitaries, it can be realized with shallow, fixed‐depth, non-parameter quantum circuits on NISQ hardware. Moreover, its grounding in evidence theory ensures that the fused state reflects coherent common support across parties, endowing the aggregation step with clear semantic interpretability.

Although the classical CCR method requires exponential complexity, the eviQVFL framework, utilizing the quantum equivalence of evidence theory, is capable of implementing CCR by only applying $K$-controlled $X$ gate for $C$ times when fusing $K$ pieces of evidence under the frame of discernment of $C$ elements \cite{luo_attribute_2024}. The $K$-controlled gate, which acts control on the first $K$ qubit and acts on the final qubit as the target, is defined by mapping the ground state:
\begin{equation}
	\ket{x_1x_2\dots x_K}\ket{r} \rightarrow \ket{x_1x_2\dots x_K}\Big|r\oplus \prod_{k=1}^Kx_k\Big\rangle,
\end{equation}
where $\oplus$ denotes modulo 2 addition. More briefly, the operation flips the last qubit only when all the first $K$ qubits are in state $\ket{1}$. Such multi-controlled X gates can be effectively implemented in the NISQ era with high fidelity \cite{arsoski2024implementing}. 

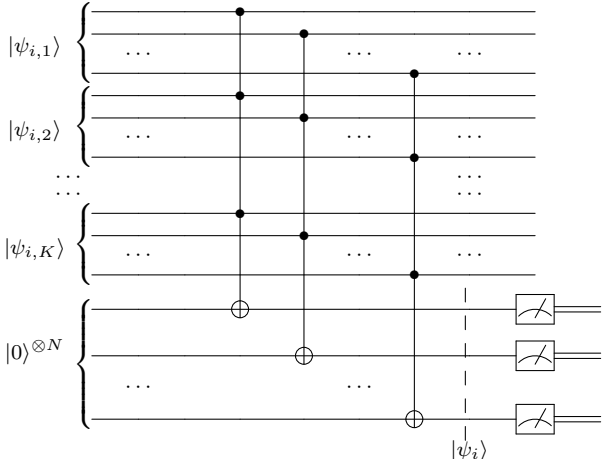
\begin{figure}[htbp]
	\centering
	\footnotesize
	\mbox{\Qcircuit @C=2.2em @R=0.8em {
			\lstick{} & \qw & \qw & \ctrl{4} & \qw & \qw & \qw & \qw & \qw\\
			\lstick{} & \qw & \qw & \qw & \ctrl{4} & \qw & \qw & \qw & \qw\\
			\lstick{} & \cdots &  &  &  & \cdots &  & \cdots & \\
			\lstick{} & \qw & \qw & \qw & \qw & \qw & \ctrl{4} & \qw & \qw
			\inputgroupv{1}{4}{.8em}{1.7em}{ \ket{\psi_{i,1}}\qquad}\\
			\lstick{} & \qw & \qw & \ctrl{6} & \qw & \qw & \qw & \qw & \qw\\
			\lstick{} & \qw & \qw & \qw & \ctrl{6} & \qw & \qw & \qw & \qw\\
			\lstick{} & \cdots &  &  &  & \cdots & & \cdots & \\ 
			\lstick{} & \qw & \qw & \qw & \qw & \qw & \ctrl{6} & \qw & \qw
			\inputgroupv{5}{8}{.8em}{1.7em}{ \ket{\psi_{i,2}}\qquad}\\
			\lstick{\cdots} &  &  &&&&& \cdots &\\
			\lstick{\cdots} &  &  &&&&& \cdots &\\
			\lstick{} & \qw & \qw & \ctrl{4} & \qw & \qw & \qw & \qw & \qw\\
			\lstick{} & \qw & \qw & \qw & \ctrl{4} & \qw & \qw & \qw & \qw\\
			\lstick{} & \cdots &  &  &  & \cdots & & \cdots & \\
			\lstick{} & \qw & \qw & \qw & \qw & \qw & \ctrl{4} & \qw & \qw
			\inputgroupv{11}{14}{.8em}{1.7em}{ \ket{\psi_{i,K}}\qquad}\\
			\lstick{} & \qw & \qw & \targ & \qw &  \qw & \qw \barrier[-0.2em]{3}&   \qw & \meter &\cw\\
			\lstick{} & \qw & \qw & \qw & \targ & \qw & \qw & \qw & \meter &\cw\\
			\lstick{} & \cdots &  &  &  & \cdots & &  & \\
			\lstick{} & \qw & \qw & \qw & \qw & \qw & \targ & \qw & \meter &\cw
			\inputgroupv{15}{18}{.8em}{2.em}{\ket{0}^{\otimes N}\qquad}\\
			&&&&&&&\mbox{$ \ket{\psi_{i}}$}&\\
		}
	}
	\caption{Illustration of a quantum circuit that fuses multiple evidence using multi-controlled $X$ gates and subsequently extracts the classical evidence by measurement after the fusion process.}
	\label{fig:evidence_fusion}
\end{figure}

The quantum circuit at the server side is shown in Fig.~\ref{fig:evidence_fusion}. The whole quantum system is composed by $K+1$ branches of qubits, which are presented as the braces in the figure. The first $K$ branches respectively carry the state teleported from $K$ server sides, and the last branch indicates the qubits for fusion result initialized as $\ket{0}^{\otimes N}$. Then, the CCR aggregating method is conducted by applying $K$-controlled $X$ gates that bring entanglement between the first $K$ brunches and the last brunch. According to the proof in \cite{luo_attribute_2024}, after gate operations, the quantum state of the last branch should be also a quantum evidence state:
\begin{equation}\label{eq:psi_final}
\begin{aligned}
      \ket{\psi_{i}}=&\sum_{x_1,\dots,x_C\in\{0,1\}}\\& \sqrt{m_{i}(F_{x_1\dots x_C})}\ \mathrm{exp}(i\beta_{x_1\dots x_C})\ \ket{x_1\dots x_C},
\end{aligned}
\end{equation}
where $ m_{i}$ denotes the evidence of the CCR aggregating result for all the evidence $m_{i,k}$ derived by distributed features at party sides, and $\beta_{x_1\dots x_C}$ can be any phase. The phase would not influence the measurement result afterwards. Essentially, each $K$-controlled X gate is controlled by qubits corresponding to the same element in the frame of discernment, realizing the  (\ref{eq:ccr}) in the quantum equivalence. 

In order to extract the output probability of the classification types from the aggregated quantum evidence state $\ket{\psi_i}$, measurements are required. Since the classification type $c$ corresponds to the $c$-th qubit, we define the projective operator as:
\begin{equation}
  \Pi_c = \bigotimes_{k=1}^{C} M_k, 
  \quad 
  M_k=\begin{cases}
    \lvert 1\rangle\!\langle 1\rvert, & k=c,\\
    I, & k\neq c.
  \end{cases}
\end{equation}
The measurement result summarizes the probability for the $c$-th qubit collapsing to $\ket{1}$. According to (\ref{eq:psi_final}), the expectation of the measurement result should be:
\begin{equation}\label{meas_out}
    \begin{aligned}
        \bra{\psi_i}\Pi_c\ket{\psi_i}&=\sum_{\substack{x_1,\dots,x_C\in\{0,1\}\\ x_c=1}} m_i(F_{x_1\dots x_C})\\
        &=\sum_{\substack{F\cap\{\omega_c\}\neq \varnothing\\F\in 2^\Omega}}m_i(F)=\mathrm{Pl}_i(\{\omega_c\})\in[0,1],
    \end{aligned}
\end{equation}
which is exactly the plausibility function $\mathrm{Pl}_i$ of the classification type hypothesis $\omega_c$. The marginal plausibility values in evidence theory is defined in (\ref{eq:plausibility_def}), determining the summation of the belief masses support $\omega_c$. The extracted classical plausibility function goes through a softmax function, deriving the final predicted label $\hat{\mathbf{y}}_i=\mathrm{softmax}\big([\mathrm{Pl}_i(\{\omega_c\})]_{c=1}^C\big)\in \mathbb{R}^C$. 

In the training process, the predicted label $\hat{\mathbf{y}}_i$ is compared to the true label ${\mathbf{y}}_i$, calculating a loss function $\mathcal{L}(\hat{\mathbf{y}}_i,{\mathbf{y}}_i)$ by the cross-entropy criterion. The local parameters $\{\boldsymbol{\theta}_k^{\mathrm{ttn}},\boldsymbol{\theta}_k^{\mathrm{vqc}}\}_{k=1}^K$ at party sides are updated by setting a classical optimizer processing the parameter gradients that obtained by the parameter-shift rules. In the practical learning process, the stochastic gradient descent (SGD) is applied, where parameters are updated by averaging the gradients of a small batch of samples. Algorithm~\ref{alg:LSB} summarizes the whole process of the eviQVFL framework.

\begin{algorithm}[htbp]
\caption{eviQVFL: Evidential Quantum Vertical Federated Learning}\label{alg:LSB}
\LinesNumbered
\KwData{Aligned samples $\{(\{\mathbf{x}_{i,k}\}_{k=1}^K,\mathbf{y}_i)\}_{i=1}^N$ with party-$k$ feature block $\mathbf{x}_{i,k}$ and server-side one-hot label $\mathbf{y}_i\in\mathbb{R}^C$.}
\KwIn{Initial party parameters $\{\boldsymbol{\theta}^{\mathrm{ttn}}_k(0),\boldsymbol{\theta}^{\mathrm{vqc}}_k(0)\}_{k=1}^K$; optimizer $\mathcal{P}$; number of rounds $T$; batch size $B$.}
\KwOut{Trained party parameters $\{\boldsymbol{\theta}^{\mathrm{ttn}}_k(T),\boldsymbol{\theta}^{\mathrm{vqc}}_k(T)\}_{k=1}^K$.}

\For{$t \leftarrow 0$ \KwTo $T-1$}{
  Sample a mini-batch $\mathcal{B}_t \subset \{1,\dots,N\}$ with $|\mathcal{B}_t|=B$\;

  \For{$i \in \mathcal{B}_t$}{
    \For{$k \leftarrow 1$ \KwTo $K$ \textbf{in parallel}}{
      Compute low-dim feature $\tilde{\mathbf{x}}_{i,k} \leftarrow \mathrm{TTN}(\mathbf{x}_{i,k};\boldsymbol{\theta}^{\mathrm{ttn}}_k(t))$\;
      Prepare encoded state $\ket{\pi_{i,k}} \leftarrow \bigotimes_{q} R_y\!\big(2\,\tilde{\mathbf{x}}_{i,k}(q)\big)\ket{0}^{\otimes n_k}$\;
      Produce local evidence state $\ket{\psi_{i,k}} \leftarrow U\!\big(\boldsymbol{\theta}^{\mathrm{vqc}}_k(t)\big)\ket{\pi_{i,k}}$ (take first $C$ qubits as output)\;
      Teleport $\ket{\psi_{i,k}}$ to the server\;
    }
    Fuse $\{\ket{\psi_{i,k}}\}_{k=1}^K$ via multi-controlled-$X$ gates to obtain $\ket{\psi_i}$\;
    Measure $\ket{\psi_i}$ with projectors $\{\Pi_c\}_{c=1}^C$ and compute $\hat{\mathbf{y}}_i=\mathrm{softmax}\big([\langle\psi_i|\Pi_c|\psi_i\rangle]_{c=1}^C\big)$\;
  }
  Compute $\mathcal{L}_t \leftarrow \frac{1}{|\mathcal{B}_t|}\sum_{i\in\mathcal{B}_t}\mathrm{CE}\!\left(\mathbf{y}_i,\hat{\mathbf{y}}_i\right)$\;
  Parameter shift to obtain $\big\{\nabla_{\boldsymbol{\theta}^{\mathrm{ttn}}_k}\mathcal{L}_t,\ \nabla_{\boldsymbol{\theta}^{\mathrm{vqc}}_k}\mathcal{L}_t\big\}_{k=1}^K$ and send gradients to parties\;

  \For{$k \leftarrow 1$ \KwTo $K$ \textbf{in parallel}}{
    $\boldsymbol{\theta}^{\mathrm{ttn}}_k(t{+}1),\,\boldsymbol{\theta}^{\mathrm{vqc}}_k(t{+}1)\leftarrow
    \mathcal{P}\!\Big(\boldsymbol{\theta}^{\mathrm{ttn}}_k(t),\boldsymbol{\theta}^{\mathrm{vqc}}_k(t);\,
    \nabla_{\boldsymbol{\theta}^{\mathrm{ttn}}_k}\mathcal{L}_t,\nabla_{\boldsymbol{\theta}^{\mathrm{vqc}}_k}\mathcal{L}_t\Big)$.\
  }
}
\end{algorithm}

\section{Experimental Results and Analysis}
\subsection{Experiment Setup}
To comprehensively evaluate the performance of our proposed eviQVFL algorithm, we introduce the following baselines for comparison.
\begin{enumerate}
    \item \textit{classical-average:} The party-side VQC in eviQVFL is replaced by a classical MLP with a single hidden layer, and the server performs aggregation by directly averaging the parties’ outputs.
    \item \textit{classical-fuse:} The party-side model is identical to classical-average; however, on the server side, the outputs from all parties are concatenated and fused via a single-layer MLP.
    \item \textit{measure-then-average:} The party-side model is identical to eviQVFL; however, each party performs a local measurement and transmits the measured outcomes to the server, which aggregates them by simple averaging.
    \item \textit{measure-then-VQC:} The party-side process is identical to measure-then-average; however, on the server side, the parties’ outputs are concatenated and encoded into a quantum circuit, processed by a two-layer VQC (with the structure shown in Fig.~\ref{fig:VQC_circuit}), and then measured.
\end{enumerate}
Additionally, we considered a model termed \textit{teleported-then-VQC}: mirrors eviQVFL in teleporting each party’s quantum state to the server, but replaces the non-trainable evidential quantum circuit with a trainable VQC for information fusion. In practice, this model consistently suffered from severe barren plateaus~\cite{mcclean2018barren,cerezo2021cost} and failed to learn even on very simple tasks—for example, binary classification in MNIST— as illustrated Fig.~\ref{fig:mnist36}. Therefore, this model is omitted from the report baseline.

In the experiments, we matched the number of trainable parameters between eviQVFL and each baseline as closely as possible to enhance comparability. The program submitted on  \url{https://github.com/luohaooo/Evidential-Vertical-Quantum-Federated-Learning} is built on \textit{torchquantum} library~\cite{hanruiwang2022quantumnas} for efficient quantum simulation using GPUs. The parameter is updated by Adam optimizer.

\subsection{Image classification}
We conduct image classification on the MNIST and FashionMNIST datasets. Each $28\times28$ grayscale image is vertically partitioned into four non‐overlapping blocks—top‐left, top‐right, bottom‐left, and bottom‐right—each of size $14\times14$. These four $196$-dimensional vectors are assigned to $K=4$ parties, so that each party $k$ receives only its local block $\xx_{i,k}\in\mathbb{R}^{196}$. 

In eviQVFL, each party’s local TTN is configured with $L=4$ TT cores, taking an input tensor of shape $[P_1,\dots P_L]=[2,7,7,2]$ and producing an output tensor of shape $[Q_1,\dots Q_L]=[1,2,2,1]$ via TT bond dimensions $\{R_l^{\cX}\}_{l=1}^{L-1},\{R_l^{\cW}\}_{l=1}^{L-1}$ set to 2. Consequently, the low-dimensional feature $\tilde{\xx}_{i,k}\in\mathbb{R}^{4}$ is obtained, which is angle-encoded onto the quantum system of $n_k=4$ qubits. The VQC module is formed as Fig.~\ref{fig:VQC_circuit} by repeating 2 dotted boxes.  For MNIST, we consider classifying digits $\{3,6\}$ and $\{2,3,5,6\}$. For FashionMNIST, the tasks are classifying $\{\text{Trouser},\text{Ankle Boot}\}$ and $\{\text{Trouser},\text{Dress},\text{Coat},\text{Ankle Boot}\}$.

\begin{figure}[htbp]
	\centering
	\begin{minipage}{0.48\linewidth}
		\centerline{\includegraphics[width=4cm]{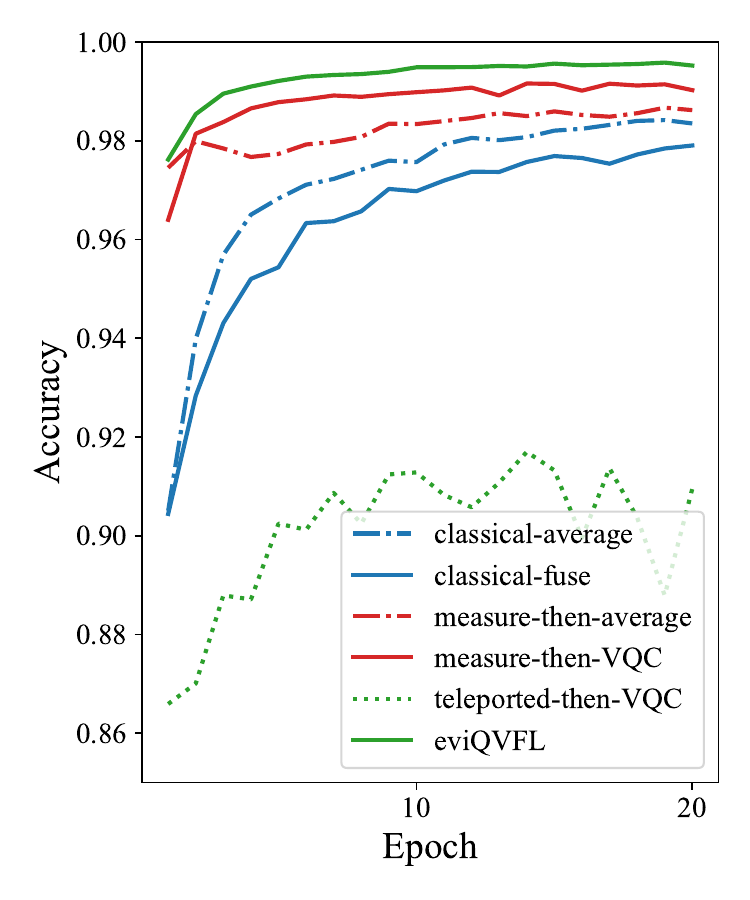}}
		\centerline{(a) Accuracy}
	\end{minipage}
	\hfill
	\begin{minipage}{.48\linewidth}
		\centerline{\includegraphics[width=4cm]{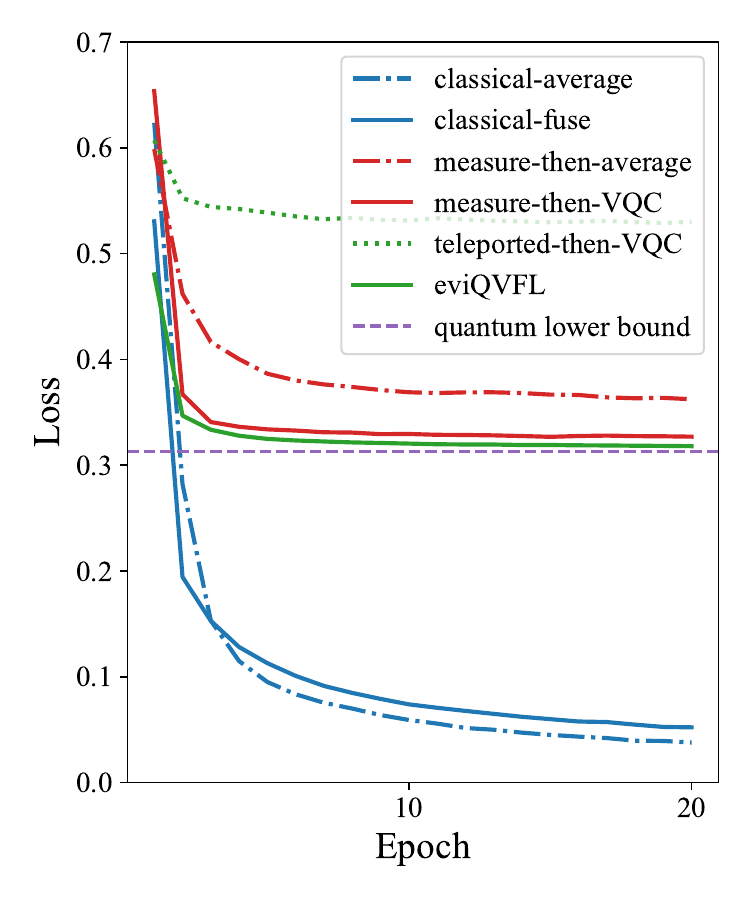}}
		\centerline{(b) Loss}
	\end{minipage}
	\caption{Classifying digits $\{3,6\}$ in MNIST dataset. }
	\label{fig:mnist36}
\end{figure}

\begin{figure}[htbp]
	\centering
	\begin{minipage}{0.48\linewidth}
		\centerline{\includegraphics[width=4cm]{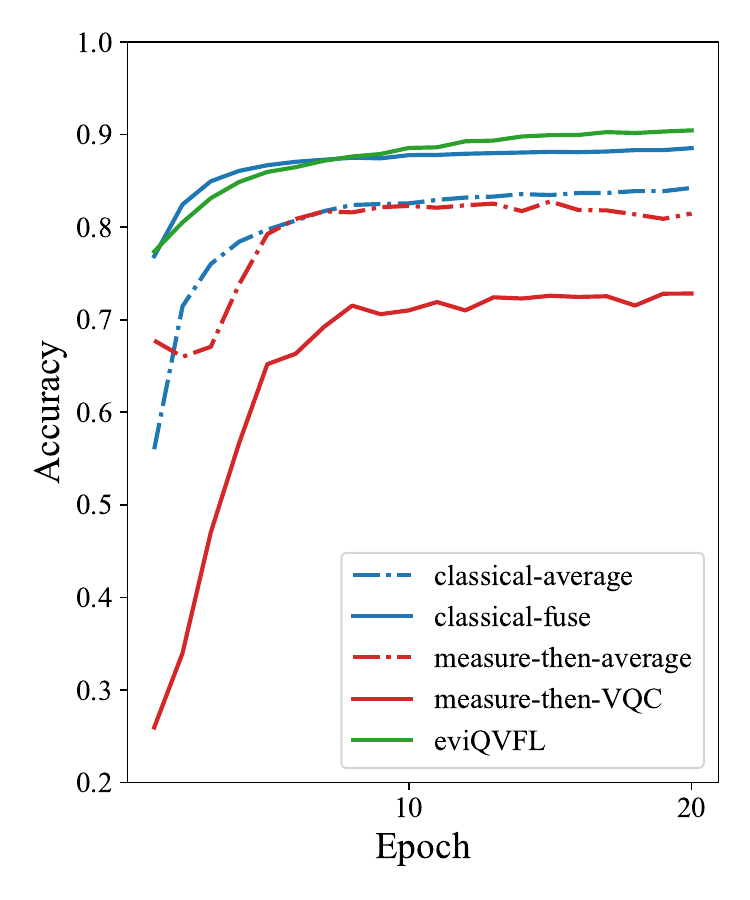}}
		\centerline{(a) Accuracy}
	\end{minipage}
	\hfill
	\begin{minipage}{.48\linewidth}
		\centerline{\includegraphics[width=4cm]{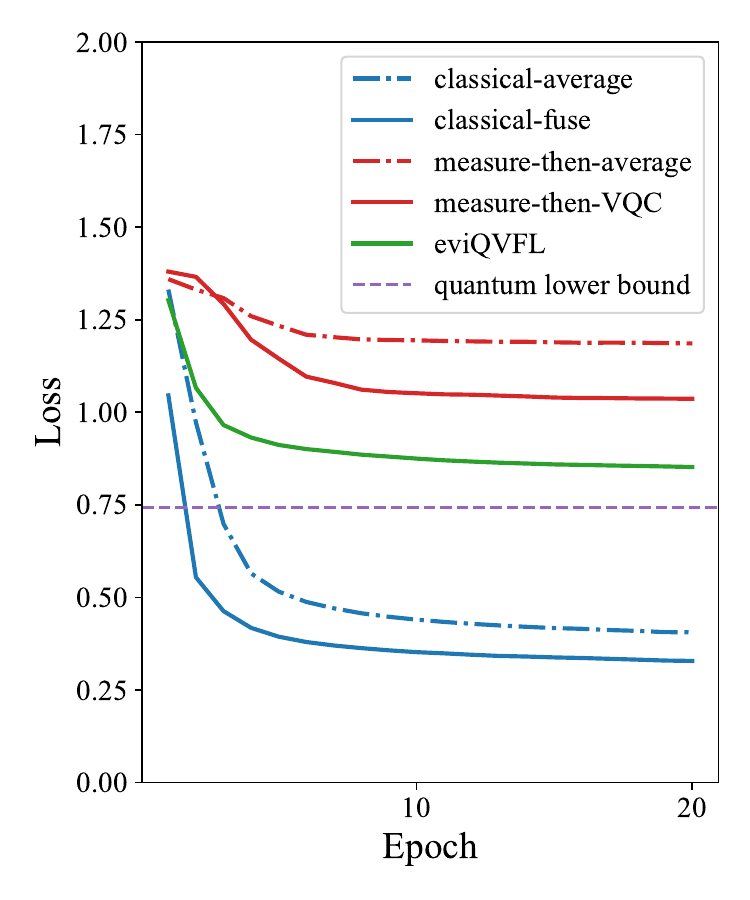}}
		\centerline{(b) Loss}
	\end{minipage}
	\caption{Classifying digits $\{2,3,5,6\}$ in MNIST dataset. }
	\label{fig:mnist2356}
\end{figure}

\begin{figure}[htbp]
	\centering
	\begin{minipage}{0.48\linewidth}
		\centerline{\includegraphics[width=4cm]{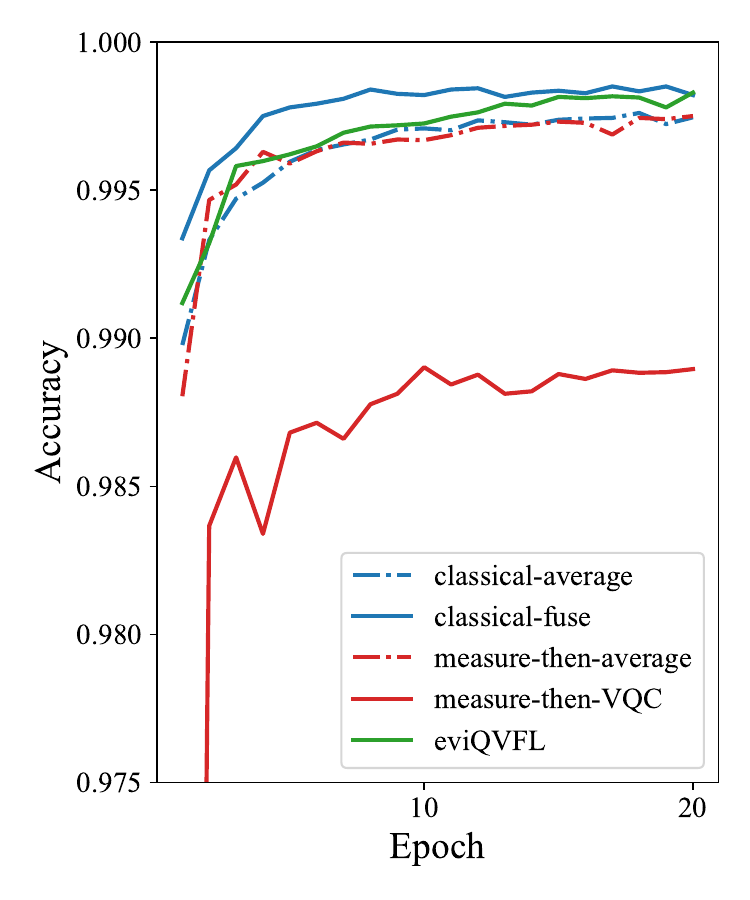}}
		\centerline{(a) Accuracy}
	\end{minipage}
	\hfill
	\begin{minipage}{.48\linewidth}
		\centerline{\includegraphics[width=4cm]{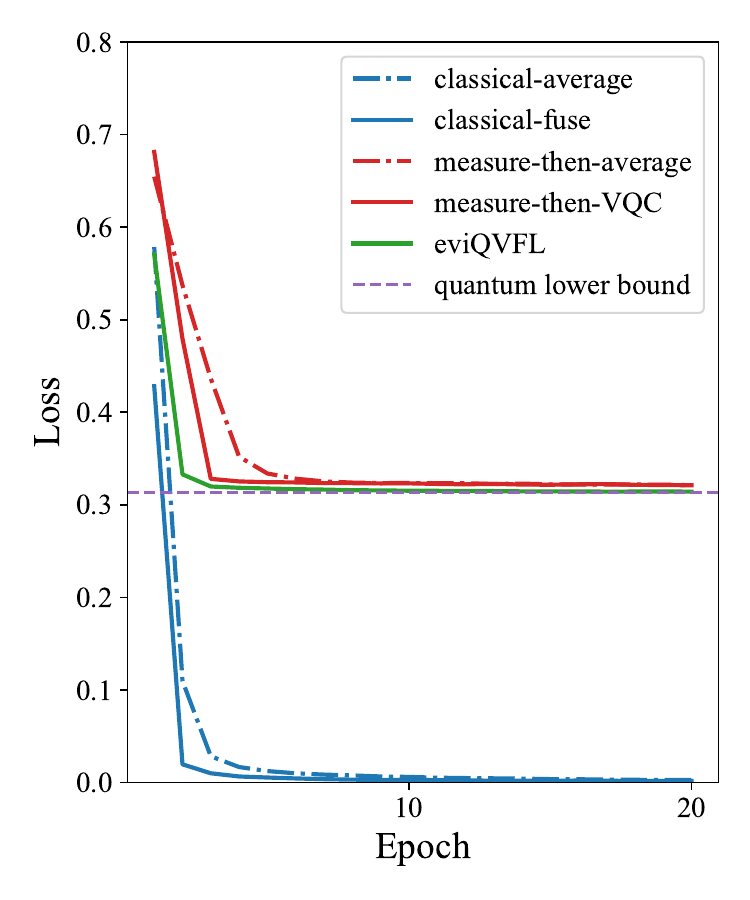}}
		\centerline{(b) Loss}
	\end{minipage}
	\caption{Classifying $\{\text{Trouser},\text{Ankle Boot}\}$ in FashionMNIST dataset. }
	\label{fig:fashion19}
\end{figure}

\begin{figure}[htbp]
	\centering
	\begin{minipage}{0.48\linewidth}
		\centerline{\includegraphics[width=4cm]{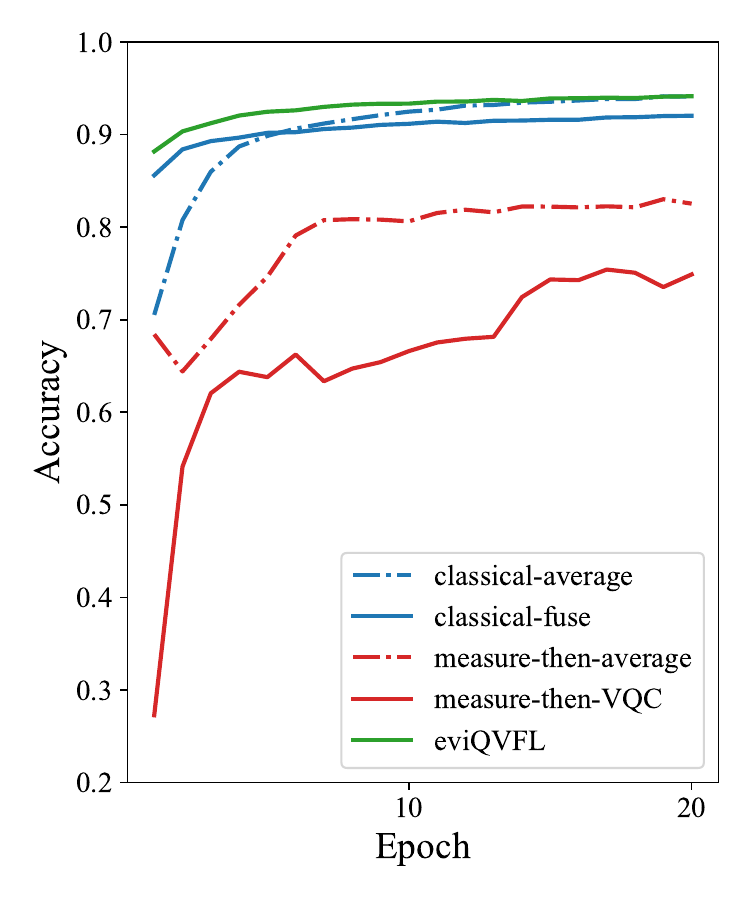}}
		\centerline{(a) Accuracy}
	\end{minipage}
	\hfill
	\begin{minipage}{.48\linewidth}
		\centerline{\includegraphics[width=4cm]{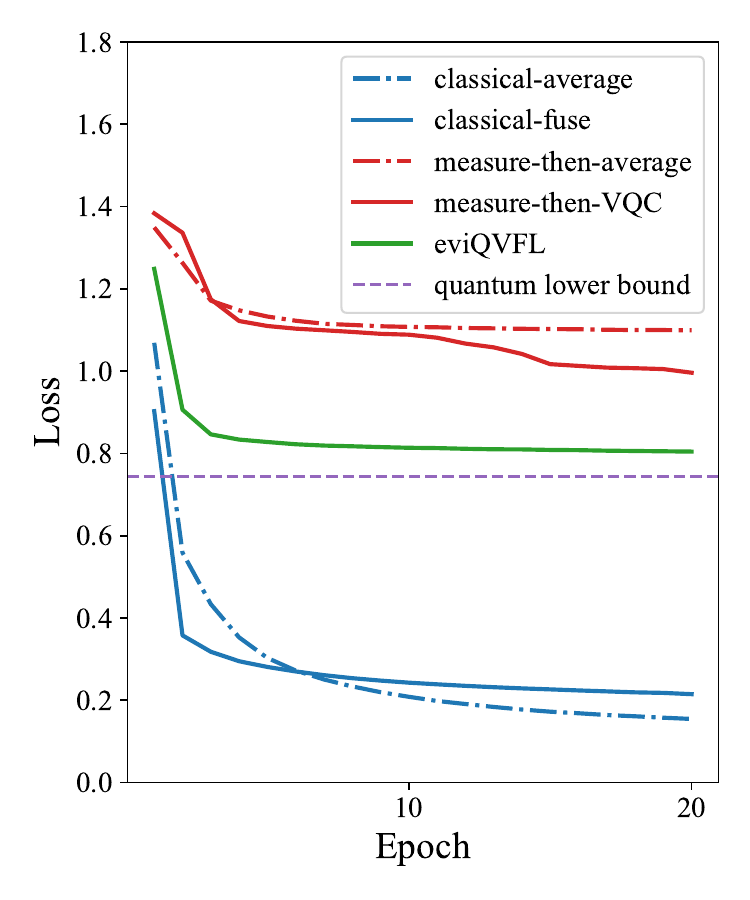}}
		\centerline{(b) Loss}
	\end{minipage}
	\caption{Classifying $\{\text{Trouser},\text{Dress},\text{Coat},\text{Ankle Boot}\}$ in FashionMNIST dataset. }
	\label{fig:fashion1349}
\end{figure}

Models are optimized using the Adam optimizer. The training set is divided into mini-batches of size 64, and each run is trained for 20 epochs. We perform 50 independent runs for each experiment and report the mean performance over these trials.  The final averaged training curves for MNIST and FashionMNIST are presented in Fig.~\ref{fig:mnist36},  Fig.~\ref{fig:mnist2356},  Fig.~\ref{fig:fashion19}, and Fig.~\ref{fig:fashion1349}, where eviQVFL stays the top accuracy among baselines.


\subsection{Breast Cancer Prediction}
\begin{figure}[htbp]
	\centering
	\begin{minipage}{0.48\linewidth}
		\centerline{\includegraphics[width=4cm]{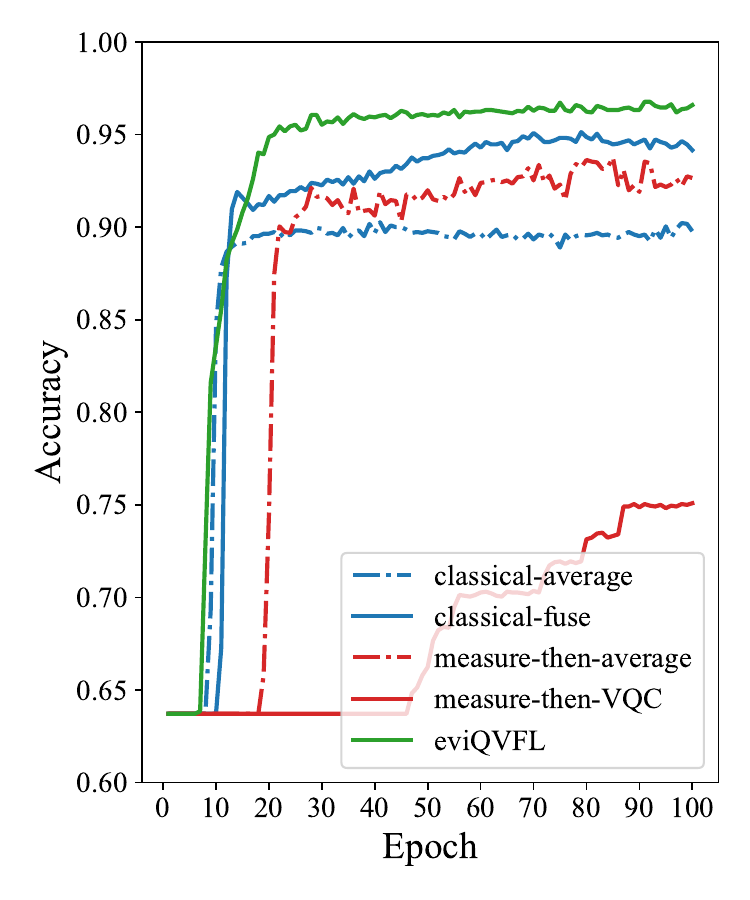}}
		\centerline{(a) Accuracy}
	\end{minipage}
	\hfill
	\begin{minipage}{.48\linewidth}
		\centerline{\includegraphics[width=4cm]{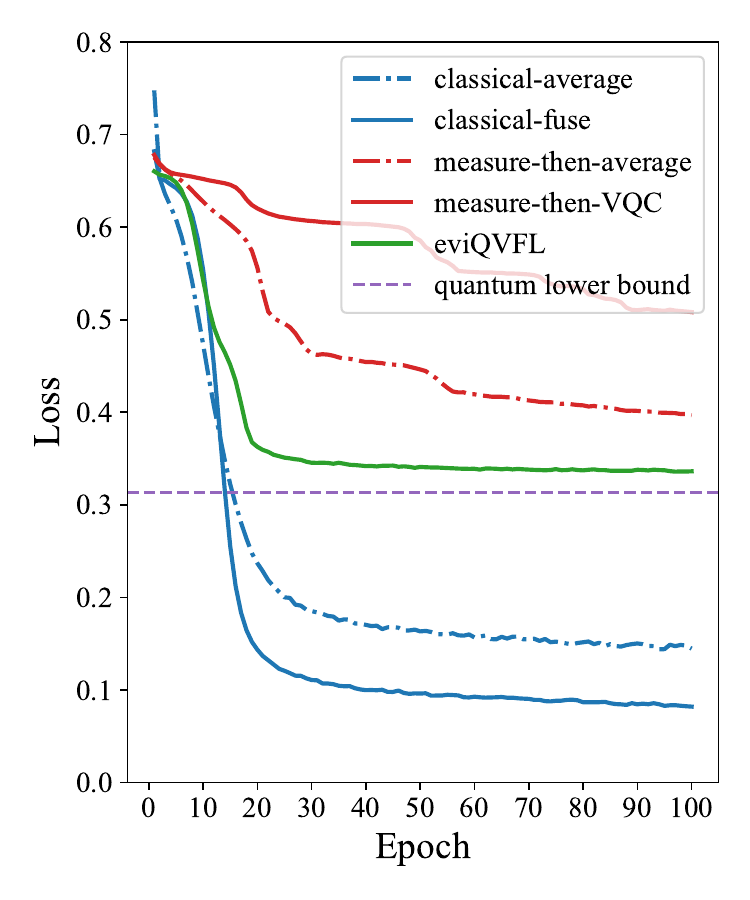}}
		\centerline{(b) Loss}
	\end{minipage}
	\caption{Predict whether the cancer is benign or malignant.}
	\label{fig:breast}
\end{figure}

Real-world VFL scenarios often involve heterogeneous, noisy tabular features. To assess eviQVFL under such conditions, we simulate algorithms with a publicly available \emph{Breast Cancer Wisconsin (Diagnostic) Dataset}, where each sample record comprises 30 real‐valued features computed from digitized images of fine needle aspirate of breast masses—and the task is to predict whether a tumor is benign or malignant ($C=2$). The features are vertically partitioned into local features $\xx_{i,k}\in\mathbb{R}^{10}$ distributed among $K=3$ parties. To reduce dimension, TT cores are configured as $[P_1,\dots P_L]=[2,5]$, $[Q_1,\dots Q_L]=[2,2]$, with bond dimension as 2. The low-dimensional feature $\tilde{\xx}_{i,k}\in\mathbb{R}^{4}$ is fed to the local quantum system of $n_k$=4 qubits. The VQC only repeats the dotted boxes 1 time. The independent run is performed 50 times to obtain the average result, which is presented in Fig.~\ref{fig:breast}. The result reveals the superior capacity of eviQVFL in the real-world dataset.

\subsection{Credit Card Fraud Detection}
\begin{figure}[htbp]
	\centering
	\begin{minipage}{0.48\linewidth}
		\centerline{\includegraphics[width=4cm]{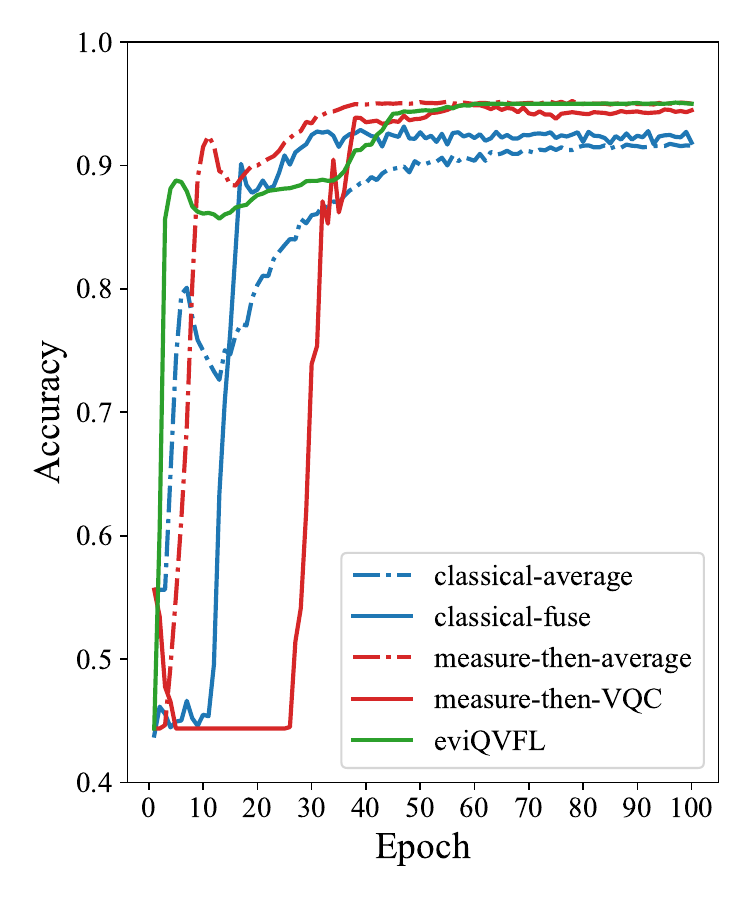}}
		\centerline{(a) Accuracy}
	\end{minipage}
	\hfill
	\begin{minipage}{.48\linewidth}
		\centerline{\includegraphics[width=4cm]{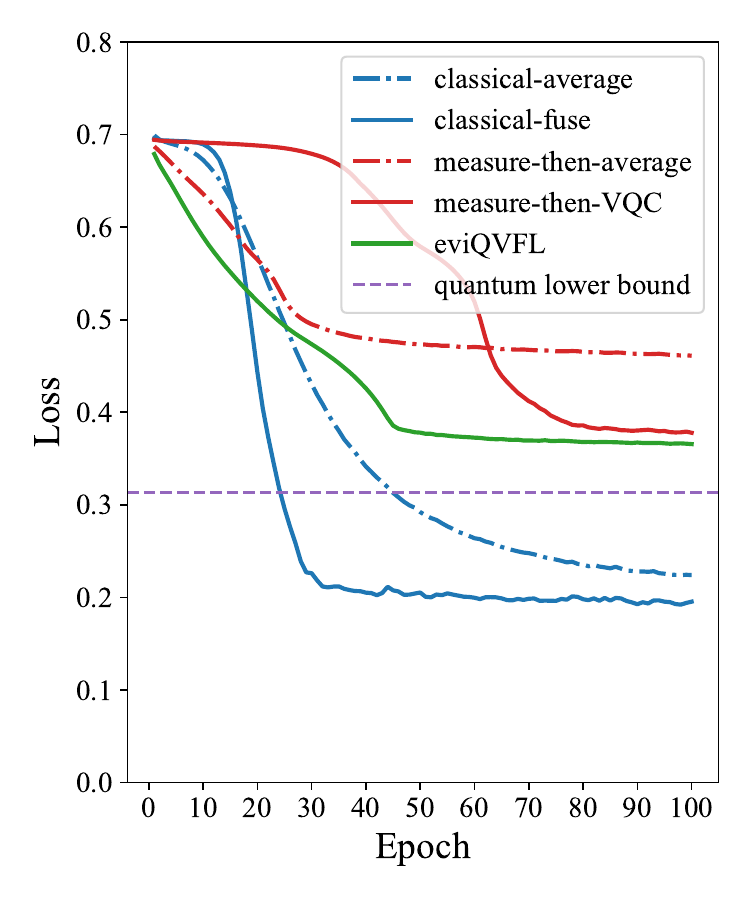}}
		\centerline{(b) Loss}
	\end{minipage}
	\caption{Detect whether the credit card transactions are fraudulent or genuine.}
	\label{fig:cf}
\end{figure}

\emph{Credit Card Fraud Detection Dataset} is another publicly available dataset, containing anonymized credit card transactions with 28 principal component features and a binary label indicating genuine or fraudulent behavior. Set $K=4$ parties, each of which holds the local feature $\mathbf{x}_{i,k}\in\mathbb{R}^{7}$. We only use one TT core $P_1=7, Q_1=3$ to transform features into $\tilde{\xx}_{i,k}\in\mathbb{R}^{3}$, where the current TTN is reduced to a single layer MLP. The VQC only repeats the dotted boxes 1 time. We perform 50 independent runs and report the averaged performance in Fig.~\ref{fig:cf}. The accuracy of eviQVFL also ranks highest.

\subsection{Analysis}
The experimental results are compiled in the table below, which also reports the parameter scale of each model. 
\begin{table}[htbp]
	\centering
	\caption{Experimental results. Report the loss, accuracy at convergence, and the number of model parameters formatted as party-side-parameter-count × number-of-parties (+ server-side-parameter-count).}
	\begin{tabular}{cccccc}
		\toprule
		Tasks & $\mathrm{C}_{\mathrm{avg}}$ & $\mathrm{C}_{\mathrm{fuse}}$ & $\mathrm{Q}_{\mathrm{avg}}$ & $\mathrm{Q}_{\mathrm{VQC}}$ & eviQVFL\\
		\hline
        \makecell{MNIST\\2-class} &\makecell{0.0379\\$98.35\%$\\144$\times$4} &\makecell{0.0522\\$97.90\%$\\144$\times$4+16} &\makecell{0.3623\\$98.62\%$\\144$\times$4} &\makecell{0.3270\\$99.02\%$\\144$\times$4+48}& \makecell{0.3180\\$99.52\%$\\144$\times$4}\\
        \hline
        \makecell{MNIST\\4-class} &\makecell{0.4057\\$84.24\%$\\152$\times$4} &\makecell{0.3285\\$88.54\%$\\152$\times$4+64} &\makecell{1.1860\\$81.74\%$\\144$\times$4} &\makecell{1.0366\\$72.83\%$\\144$\times$4+48}& \makecell{0.8522\\$90.45\%$\\144$\times$4}\\
        \hline
        \makecell{Fashion\\-MNIST\\2-class} &\makecell{0.0028\\$99.75\%$\\144$\times$4} &\makecell{0.0018\\$99.82\%$\\144$\times$4+64} &\makecell{0.3214\\$99.75\%$\\144$\times$4} &\makecell{0.3212\\$98.90\%$\\144$\times$4+48}& \makecell{0.3143\\$99.83\%$\\144$\times$4}\\
        \hline
        \makecell{Fashion\\-MNIST\\4-class}&\makecell{0.1546\\$94.09\%$\\152$\times$4} &\makecell{0.2149\\$92.03\%$\\152$\times$4+64} &\makecell{1.0992\\$82.54\%$\\144$\times$4}  &\makecell{0.9962\\$74.89\%$\\144$\times$4+48}& \makecell{0.8045\\$94.14\%$\\144$\times$4}\\
        \hline
        \makecell{Breast\\Cancer\\2-class}&\makecell{0.1447\\$89.78\%$\\40$\times$3} &\makecell{0.0821\\$94.16\%$\\40$\times$3+12} &\makecell{0.3970\\$92.65\%$\\40$\times$3} &\makecell{0.5082\\$75.09\%$\\40$\times$3+36}& \makecell{0.3362\\$96.59\%$\\40$\times$3}\\
        \hline
        \makecell{Credit\\Card\\2-class}&\makecell{0.2242\\$91.59\%$\\31$\times$4} &\makecell{0.1955\\$91.84\%$\\31$\times$4+16} &\makecell{0.4611\\$95.00\%$\\30$\times$4} &\makecell{0.3779\\$94.47\%$\\30$\times$4+48}& \makecell{0.3656\\$95.00\%$\\30$\times$4}\\
		\bottomrule
	\end{tabular}
	\label{tab:results}
\end{table}

Inspecting these results, we find that eviQVFL achieves higher accuracy than all baselines across every test; however, the losses of the classical models are uniformly lower than those of the quantum models. This seemingly counterintuitive pattern stems from the fact that quantum measurement outputs, as described in (\ref{meas_out}), are strictly confined to $[0,1]$, which induces a constant lower bound for the cross-entropy (CE) loss marked as the purple dashed line in the figures.

\prop \label{lm1}
In a $C$ class classification task, the CE loss for models with the quantum measurement as the output has a lower bound: $\mathcal{L}\geq\ln(C+e-1)-1$.

\emph{Proof.\ } Let $\hat{\mathbf{y}}=[\hat{y}_i(1),\ldots,\hat{y}_i(C)]$ with $\hat{y}_i(c)\in [0,1]$ be the output logits. Without loss of generality, assume the correct class is the class 1. Thus, the loss satisfies:
\begin{equation}
	\begin{aligned}
			\mathcal{L}&=-\ln\frac{e^{\hat{y}_i(1)}}{\sum_{c=1}^Ce^{\hat{y}_i(c)}}=\ln\left( 1+\frac{\sum_{c=2}^Ce^{\hat{y}_i(c)}}{{e^{\hat{y}_i(1)}}}	\right)\\
			&\geq\ln\left(1+\frac{C-1}{e}	\right)=\ln(C+e-1)-1. 
	\end{aligned}
\end{equation}
\hfill $\Box$

However, the classical models output unbounded real-valued logits, for which the loss is only bounded below by zero. Thus, the losses of quantum and classical models are not directly comparable. Nevertheless, the experimental results show that, under the same parameter budget, eviQVFL always achieves higher accuracy, highlighting its stronger representational capacity.

Among quantum models, the losses are directly comparable. The results show that eviQVFL consistently attains lower loss and achieves higher accuracy, whether the baseline employs a parameter-free server or a parameterized server. Since all methods share identical party-side submodels, these gains highlight the advantages of quantum teleportation and the evidential fusion method in eviQVFL. In detail, eviQVFL and the \textit{quantum-then-average} baseline employ the same party-side TTN-VQC architecture, and both are non-trainable on the server side. We can analyze their respective representational capacities as follows. We impose a continuity assumption on the loss function.

\assump\label{as}
 Loss functions $\cL(\yy_i,\cdot)$ are Lipschitz continuous with parameter $\omega$, i.e. for any $\zz_1,\zz_2 \in \mathbb{R}^{C}$,
\begin{align}
    \norm{\cL(\yy_i,\zz_1)-\cL(\yy_i,\zz_2)}_1\leq \omega \norm{\zz_1-\zz_2}_1.
\end{align}
Therefore, the loss can be bounded by:
\begin{align}
    \cL(\yy_i,\hat{\yy}_i)\leq \omega\norm{\yy_i-\hat{\yy}_i}_1.
\end{align}
To evaluate the representation capacity, the approximation error can be used as the metric, which measures the error in approximating a target function when applying the optimal parameters for the given model families.
In \cite{qi2023theoretical}, Qi \emph{et al.} derive the approximation error bound of the TTN-VQC model in the context of functional regression. According to Theorem 1 in \cite{qi2023theoretical}, the following bound is introduced.
\prop \label{lm2}
Suggest a functional regression problem that maps an input vector space $\mathbb{R}^Q$ to an output vector space $\mathbb{R}^U$. Then, under the mean absolute error (MAE) loss $\cL$, given a data distribution $\cD$ for the input, assuming a smooth target function $h_{\cD}^*$, there exists an optimal TTN-VQC operator $f_{\cD}^*$ that guarantees the expected loss:
\begin{equation}
\begin{aligned}
    \mathbb{E}(\cL)&=\mathbb{E}_{\xx\sim \cD}\left(\norm{h_{\cD}^*(\xx)-f_{\cD}^*(\xx)}_1\right)\\
    &\leq \frac{\Theta(1)}{\sqrt{U}} + \cO\left(\frac{1}{\sqrt{M}}\right),
\end{aligned}
\end{equation}
where $U$ and $M$ separately refer to the number of qubits and the count of quantum measurements. The bound implies that the approximation error decays as $U$ and $M$ increase. 

\emph{Proof.\ } The approximation error between $h_{\cD}^*$ and $f_{\cD}^*$ is composed by the algorithm error term caused by the algorithm's finite representation power and a sampling error term caused by the finite measurements of the quantum model. Applying the triangle inequality to $\norm{h_{\cD}^*(\xx)-f_{\cD}^*(\xx)}_1$ yields an upper bound equal to the sum of the $\ell_1$-norms of these two error terms. TTN-VQC structure can be viewed as a convex hull of $U$ Sigmoid activation function. According to Barron’s universal approximation theorem~\cite{barron1994approximation}, the former term has the upper bound of $\Theta(1)/\sqrt{U}$. The latter term can be controlled via the Central Limit Theorem: for i.i.d. $M$ samples with finite variance, the sampling error decays at the rate of $\cO(1/M)$.\hfill $\Box$

View eviQVFL from a global perspective, the party-side TTN-VQCs and the evidential fusion circuit, linked via quantum teleportation, together form an effective TTN-VQC of $\sum_{k=1}^K n_k$ qubits in total. The only difference is that the cross-party TT bond is specially fixed to zero in this scenario. We denote $\cD_k$ as the distribution for the $k$-th feature $\xx_{i,k}$, and $\cD$ as the joint distribution for all the features $\{\xx_{i,k}\}_{K=1}^K$. 
According to Assumption~\ref{as} and Proposition~\ref{lm2}, the approximation error of the joint model admits an upper bound:
\begin{equation}
	\begin{aligned}
		    \mathbb{E}(\cL)&\leq \omega\left( \mathbb{E}_{\{\xx_{i,k}\}_{K=1}^K\sim\cD}\left( \norm{\yy_i-\hat{\yy}_i}_1 \right)\right)\\
		&\leq \frac{\Theta(1)}{\sqrt{\sum_{k=1}^K n_k}} + \cO(\frac{1}{\sqrt{M}}).\label{eq:eviqvfl}
	\end{aligned}
\end{equation}
In contrast, for the \textit{quantum-then-average} baseline, the approximation error can be derived by triangle inequality:
\begin{equation}
	\begin{aligned}
		   \mathbb{E}(\cL)&\leq \omega \left( \mathbb{E}_{\{\xx_{i,k}\}_{K=1}^K\sim\cD}\left( \norm{\yy_i-\frac{1}{K}\sum_{k=1}^K\hat{\yy}_{i,k}}_1 \right)   \right)\\
		&\leq\omega\left( \frac{1}{K} \sum_{k=1}^K \mathbb{E}_{\xx_{i,k}\sim\cD_k}\left( \norm{\yy_i-\hat{\yy}_{i,k}}_1 \right)  \right)\\
		&\leq  \frac{1}{K}\sum_{k=1}^K\frac{1}{\sqrt{n_k}}\Theta(1)+ \cO(\frac{1}{\sqrt{M}}).\label{eq:avgqvfl}
	\end{aligned}
\end{equation}

Comparing the bounds in (\ref{eq:eviqvfl}) and  (\ref{eq:avgqvfl}), the approximation error of the eviQVFL joint model decays with the aggregate qubit budget for all the parties, which guarantees the representation capacity of that all the available quantum resources could provide. However,  the \textit{quantum-then-average} baseline only 
guarantees the average representation capacity of local models. 

On the other hand, although the evidential fusion circuit in eviQVFL is non-trainable, it is resilient to the notorious barren plateaus phenomenon in QML, that is when training a VQC the expected gradient approaches zero and its variance decays exponentially with the number of qubits involved, leaving virtually no usable learning signal and causing training to stagnate. McClean \textit{et al.} provide a detailed analysis and rigorous proofs of this phenomenon~\cite{mcclean2018barren,cerezo2021cost}. If one uses random initialization and unstructured parameterized quantum circuits to explore a Hilbert space, increasing the number of qubits and the circuit depth typically exacerbates barren plateaus. For example, the previously noted \textit{teleported-then-VQC} model that is omitted from the baselines utilize a VQC of $\sum_{k=1}^Kn_k$ qubits is nearly untrainable. One remedy is to employ parameterized, task-informed structured circuits, i.e., architectures whose design is guided by insights from the problem itself. For the VFL classification setting, our eviQVFL instantiates this idea by drawing on the information-fusion and decision-making principles of evidence theory: We use quantum teleportation together with an evidential fusion circuit to aggregate party-side features and produce class predictions, rather than directly utilize a random VQC. Despit the party-side VQC circuits are still random selected, our experiments indicate that this design effectively alleviates the barren plateau phenomenon. As shown in figures, eviQVFL is always trainable and converges faster than the \textit{quantum-then-VQC} baseline. The contrast is most pronounced in Fig.~\ref{fig:breast} and Fig.~\ref{fig:cf}, where \textit{quantum-then-VQC} encounters severe barren plateaus, exhibiting virtually no loss reduction in the initial epochs. Note that the reported curves are averages over 50 independent runs.

Beyond the superior performance and trainability, a notable advantage over the other baselines is that eviQVFL inherently preserves the privacy of party-side output by design. Since party-side transmits the local quantum state output $\ket{\psi_{i,k}}$  to the server by quantum teleportation, the server cannot directly obtain classical observations of the parties’ output states due to the no-cloning theorem and the measurement postulate that collapses quantum states upon observation. This intrinsic property of quantum mechanics provides stronger privacy protection, whereas classical VFL frameworks typically require additional techniques such as homomorphic encryption to achieve such security~\cite{hardy2017private, cheng2021secureboost}.

\section{Conclusion}
We have introduced eviQVFL, the first QFL framework tailored to the VFL setting. The eviQVFL framework is composed by party-side TTN-VQC model for local feature processing and server-side quantum circuit for fusing party-side output states, which is transmitted via quantum teleportation. We adopt a non-parametric fusion scheme on the server-side, implemented by an evidential fusion circuit informed by evidence theory for multi-source information fusion. In experiments including image classification, cancer diagnosis, and credit‐card fraud detection, we find that eviQVFL consistently achieves higher classification accuracy than both classical and quantum baselines, and attains lower loss relative to other quantum baselines. The eviQVFL framework employs quantum teleportation to transmit information from the parties to the server, thereby endowing the model with a representational capacity determined by the total quantum resources available across all parties, rather than being constrained by any single party’s local subset of resources. Besides, our results indicate that a fixed evidential fusion layer substantially alleviates barren plateaus during VQC training. 

The eviQVFL framework provides a quantum template model for addressing VFL problem. Future work will investigate alternative party-side fusion schemes and extend the framework to handle multimodal data, including text, images, and graph data, to accommodate diverse use cases. Besides, eviQVFL constitutes an effective integration of evidence theory with QML, demonstrating how insights from evidence theory can inform the design of interpretable and trainable VQC. A foreseeable direction is to bind richer evidence-theoretic update mechanisms to quantum circuits, thereby enhancing the capability of VQCs.

\bibliographystyle{IEEEtran}
\bibliography{bibtex}

\end{document}